\documentclass[preprint,12pt,authoryear]{elsarticle}
\usepackage{lineno,hyperref}
\usepackage{ragged2e}
\usepackage[margin=1in]{geometry}
\usepackage{graphicx,subfigure}
\usepackage{multirow}
\usepackage{epstopdf}
\usepackage{setspace}
\usepackage{scalerel}
\usepackage{bbold}
\usepackage{float}
\usepackage{amsmath, amssymb, ulem}
\usepackage{rotating}
\usepackage{morefloats}
\usepackage{stackengine}
\usepackage{comment}
\usepackage{caption}
\usepackage{natbib}
\setcitestyle{round}
\usepackage{fancybox,color}
\usepackage{booktabs}
\usepackage[utf8]{inputenc}
\usepackage[english]{babel}

\usepackage{multirow}

\usepackage{adjustbox}
\usepackage{amsmath,amssymb,amsthm}
\usepackage{thmtools}
\declaretheoremstyle[
spaceabove=6pt, spacebelow=6pt,
headfont=\normalfont\bfseries,
notefont=\mdseries, notebraces={(}{)},
bodyfont=\normalfont,
postheadspace=0.6em,
headpunct=:
]{mystyle}

\usepackage{graphicx}
\usepackage{cleveref}
\crefname{hyp}{hypothesis}{hypotheses}
\Crefname{hyp}{Hypothesis}{Hypotheses}
\usepackage{rotating,tikz}
\usepackage{subfigure}
\newcommand{\mathz}{\ooalign{$Z$\cr\hfil\rule[.8ex]{.4em}{.08ex}\hfil\cr}}

\usepackage{xcolor}

\usepackage{subcaption}

\makeatletter 
\renewcommand\@biblabel[1]{} 
\makeatother

\journal{Expert Systems with Applications}

\usepackage[margin=1in]{geometry}
\usepackage{amsmath,amsthm,amssymb,amsfonts,graphicx,dsfont}

\usepackage{booktabs,parskip}
\usepackage{epsfig}

\usepackage{url}
\usepackage{bm}
\usepackage{rotating}
\usepackage{listings}
\usepackage{verbatim}
\usepackage{xcolor}
\usepackage[title]{appendix}
\usepackage{multirow}

\usepackage[ruled,vlined,linesnumbered]{algorithm2e}
\SetKwFunction{KwFn}{Fn}
\SetKwInOut{Input}{input}
\SetKwInOut{Output}{output}

\usepackage{longtable}
\usepackage{adjustbox}
\usepackage{subfigure}
\hypersetup{colorlinks,%
citecolor=black,%
filecolor=black,%
linkcolor=black,%
urlcolor=black
}

\usepackage{setspace} 

\newcommand{\eval}[2][\right]{\relax
	\ifx#1\right\relax \left.\fi#2#1\rvert}

\newtheoremstyle{break}
  {\topsep}{\topsep}%
  {\itshape}{}%
  {\bfseries}{}%
  {\newline}{}%
\theoremstyle{break}

\newtheorem{remark}{Remark}

\newtheorem{proposition}{Proposition}
\theoremstyle{definition}

\newtheorem{exampleemph}[proposition]{Example}   

\usepackage{changepage}   
\makeatletter
\newcommand*{\rom}[1]{\expandafter\@slowromancap\romannumeral #1@}
\makeatother

\journal{Expert Systems with Applications}

\begin{document}

\begin{frontmatter}

\title{Topological Clustering of Agents in Information Contagions: Application to Financial Markets
}
-
\author[mymainaddress]{Anubha Goel\corref{mycorrespondingauthor}}
\cortext[mycorrespondingauthor]{Corresponding author. Tel.: +358 503201908.}
\ead{anubha.goel@tuni.fi}		
		\author[mymainaddress]{Henri Hansen}
		\ead{henri.hensen@tuni.fi} 
		
		\author[mymainaddress]{Juho Kanniainen}
		\ead{juho.kanniainen@tuni.fi}
		
		\address[mymainaddress]{Computing Science/Financial Computing and Data Analytics Group, Tampere University, Tampere, 33720, Finland}

\begin{abstract}

Building on topological data analysis and expert knowledge, this study introduces a Mapper-based approach to cluster agents based on their tendency to be influenced by information spread. The context of our paper is financial markets with an aim to identify agents trading opportunistically on insider information while minimizing false positives, a critical challenge in financial market surveillance. We verify and demonstrate our methods using both synthetic and empirical data on insider networks and investor-level transactions in a stock market. Recognizing the sensitive nature of insider trading cases, we design a conservative approach to minimize false positives, ensuring that innocent agents are not wrongfully implicated. We find that the mapper-based method systematically outperforms other methods on synthetic data with ground truth. We also apply the method to empirical data and verify the results using a statistical validation method based on persistence homology. Our findings indicate that the proposed Mapper-based technique effectively identifies a subset of agents who tend to take advantage of inside information they have received. This method is highly adaptable to various applications involving the spread of information or diseases, where agents exhibit only indirect evidence of their carrier status (symptoms) through their behavior.

\end{abstract}

\begin{highlights}

    \item \textbf{Novel Mapper-Based Topological Data Analysis Approach}: 
    Proposes a Mapper-based approach with expert knowledge to identify agents likely to be exposed to and influenced by information spread while minimizing false positives.
    \item \textbf{Validated on Synthetic \& Empirical Data}: Validates the approach using synthetic data with ground truth and empirical data with the statistical analysis using topological features of the data.
    \item \textbf{Superior Performance in Market Surveillance}: Outperforms traditional anomaly detection and clustering methods in identifying information-driven trading behaviors.
    \item \textbf{Broad Applicability}: Extends beyond finance to disease spreading and misinformation analysis, offering a versatile tool for contagion detection.

\end{highlights}


\end{frontmatter}



\section{Introduction}

An information contagion occurs when individuals, referred to here as \emph{agents}, in a social network receive information from their neighbors and are influenced by them. These individuals in turn then influence their neighbors and so forth, resulting in a cascade. Although information contagions can occur in various domains, they are particularly significant in financial markets, where investors seek to leverage private information. Imagine a scenario where insider information about an upcoming public announcement spreads within a social network before the official release. Those who receive this valuable insider information early can take an advantage of it and trade profitably by buying or selling stocks in anticipation of the expected market movement. While these trades are not necessarily public and may not be immediately noticed by other investors, individuals with insider information may share it further with their social contacts, spreading the information within the network. When many people gain access to this insider knowledge, and some of them engage in profitable trading activities and spread information further, a cascade effect is created. This influences a larger group to trade similarly, which amplifies the market impact even before the official announcement.

Market supervisors can rarely observe who is taking advantage of private information and who is not. Instead, methods relying on indirect evidence from agents’ actions in stock markets are required. However, this is methodologically a challenging task, as trading activities are influenced not only by the private information in question but also by various other factors, such as public information, investors’ interpretations, and trading schedules.
That is, an agent making profitable trades before public announcements about a company may appear suspicious, yet similar behavior could also result from unknown factors or mere chance.

Methodologically, standard statistical and machine learning methods face challenges when no ground truth about true activation status is known and there are limited observations for each agent. Moreover, the data can be not only sparse but also multidimensional, for which conventional unsupervised clustering methods are not necessarily efficient. In this paper, we propose a method based on Topological Data Analysis (TDA), specifically the Mapper Algorithm to address this problem. Our approach uses expert knowledge to detect investor clusters with interesting topological structures that traditional methods often fail to identify. The main advantage of Mapper is that it works as a soft clustering technique which makes use of expert knowledge to form filter functions, which are then used in data projection. It extracts simple descriptions of original data in the form of simplicial complexes.

We apply the method to profile agents in financial markets based on their trading patterns, identifying those who seem to receive inside information before it is publicly announced. We refer to such agents as ``opportunistic'' and our aim is to detect these agents as accurately as possible.   Our emphasis is on reducing false positives while still achieving a substantial number of true positive identifications, and demonstrate that the Mapper technique is ideally suited for this purpose. While the application of this paper focuses on financial markets, the method is highly generalizable and applicable to situations where the goal is to identify infected individuals. The requirements include indirect observations of agents’ behavior over multiple episodes, and it is beneficial to have at least partial observations of social connections between the agents.

Because no ground truth about agent type is known empirically, we first create a Monte Carlo experiment to validate our method before applying it to real-world data. These validation results from Monte-Carlo experiment are reported in \ref{app_monte}. We then apply the Mapper-based method to a unique empirical trading data from actual investors who are a part of the social network of company insiders.  The method aims to identify agents who opportunistically takes an advantage of private information for each company separately, analyzing all investor-company pairs. For $N$ investors investing across $M$ companies, this results in $M\times N$ investor company pairs to be clustered. 

In addition to investor-level transaction data, we have a (partially) observable social network based on joint insider positions and family ties. However, since this type of information is rarely available, using the insider network for clustering would make our method dependent on data that is not widely accessible, limiting its applicability. Therefore, we use it to validate our results rather than incorporating it into the clustering process. We find that social links have strong association with how agents are profiled by our approach, providing compelling evidence that observable social connections mediate differences in trading behavior. To statistically validate our findings, we utilize the Gibbs distributions for persistence diagrams (GPD) to obtain the distributional forms of persistence diagrams for opportunistic agents identified by the Mapper approach and compare them to other agents, both in pre-announcement and non-announcement periods. Our analysis reveals statistically significant differences in the distributions of opportunistic agents, and also higher distances between the topological summaries during pre-announcement and announcement periods for these agents compared to others. 

Section 2 provides a literature review on information spreading in financial markets and introduces the necessary preliminaries on the Mapper algorithm and an overview of persistent homology and persistence diagrams. Section 3 outlines the proposed methodology. Section 4 applies this methodology to analyze insider information in stock markets. Section 5 presents the empirical data and key findings on insider networks, incorporating both Monte Carlo experiments and real-world data analysis. Finally, Section 6 concludes the paper by synthesizing the results and discussing their broader implications.

\section{Background}

In this section, we provide a brief literature review on information spreading in financial markets and introduce the theoretical foundations underlying the Mapper algorithm and provide essential concepts related to persistent homology and persistence diagrams, focusing specifically on superlevel-set filtration. For further details, we refer the reader to Ref. \citep{fasy2014confidence}.

\subsection{Literature on Information Spreading in Financial Markets}

In the analysis of information or epidemic contagion, reliable inference from a given model is central. Several methods have been suggested for this purpose, such as belief propagation and dynamic message-passing algorithms \citep{gokcen2022disentangling,pearl2022fusion}. However, these methods assume that the activation status of the agents, that is, whether they have received and reacted to the information, is known and observable \citep{gomez2012inferring,newman2023message,pouget2015inferring,wilinski2021prediction,lokhov2024learning}. In practice, the activation status in both information and epidemic contagions is often hidden and not directly observable. For example, during the COVID-19 pandemic, infection chains and the actual infection statuses of individuals were frequently hidden due to numerous asymptomatic cases, frequent false-positive symptoms, and inaccuracies in rapid antigen tests.\footnote{Rapid antigen tests were commonly used instead of the more accurate RT-PCR tests for quickly identifying infectious contacts, despite their potential for both false positives and negatives, which introduced uncertainty regarding individuals' actual health status. For more information on their use, see, for example, \url{https://www.ecdc.europa.eu/sites/default/files/documents/Options-use-of-rapid-antigen-tests-for-COVID-19_0.pdf}.}

In the existing literature, the question of identifying informed investors is rarely addressed, possibly due to a lack of data and reliable methods. One example is Ref. \citep{baltakys2023predicting}, which employs Graph Neural Networks to forecast investors' trading behavior based on trading within their local social networks. However, they do not incorporate information events such as announcements and focus solely on the predictability of individuals' transactions within a social network. This approach only indirectly provides evidence that private information is being utilized. There is an expanding body of research on so-called investor networks, which aims at identifying information links between investors by detecting pairs with significantly high synchronization in their trading behavior \citep{ozsoylev2014investor,tumminello2012identification,ranganathan2018dynamics,baltakiene2021identification}. This field focuses on predicting effective links between investors rather than pinpointing the most opportunistic traders based on their trading behavior. One relevant study is Ref. \citep{ahern2017information}, which analyzes legal documents on both inside traders and people who shared information but did not trade securities themselves. The study discovers that inside information typically originates from corporate executives and that people who trade on this inside information see substantial returns of approximately 35\% over a 21-day period. Unlike in Ref. \citep{ahern2017information}, in which direct observations of information arrival were used to discover and identify recipients of information, our objective is to identify whether an investor traded using private information.

\subsection{TDA Mapper}
The Mapper algorithm, first introduced by \citep{singh2007topological}, is a topological data analysis (TDA) method derived from the concept of generalized Reeb graphs. It effectively transforms complex, high-dimensional datasets into simplified graph structures, facilitating easier visualization and interpretation. Recent studies, such as \citep{brown2021probabilistic}, have addressed the theoretical stability of Mapper graphs, further establishing their reliability and robustness.

The Mapper procedure is built upon two primary components: a dataset \(X\) equipped with a metric \(d\), and a filter (or lens) function \(f: X \rightarrow \mathbb{R}^d\). The process involves the following key steps:

\textbf{Lens Function}: Initially, a carefully selected filter function \(f\) maps the high-dimensional dataset \(X\) into a lower-dimensional representation \(\mathbb{R}^d\). This projection reduces complexity while preserving meaningful data characteristics aligned with the research objectives.

\textbf{Covering and Level Sets}: The projected dataset \(f(X)\) is then covered by a finite collection of overlapping sets (level sets), \(\mathbf{U}=\{U_i\}_{i \in I}\), satisfying \(f(X) \subseteq \bigcup_{i \in I} U_i\). Importantly, this overlapping coverage groups data points based on their proximity in the filtered space, rather than direct similarity in the original dataset.

\textbf{Pullback Cover}: Subsequently, each covering set \(U_i\) in \(\mathbf{U}\) is mapped back to the original data space to create pre-images, or pullback sets, defined by \(f^{-1}(U_i)\). These subsets include data points sharing the same range of values under the filter function \(f\), though they may not necessarily be globally similar within the original space \(X\).

\textbf{Clustering}: Within each pre-image \(f^{-1}(U_i)\), we apply a clustering algorithm, utilizing the distance metric \(d\). This step yields distinct clusters \(\{C_{i_1}, C_{i_2}, \dots, C_{i_j}\}\) representing locally connected data points. Notably, clustering is performed in the original data space, allowing Mapper to simultaneously capture global structure (through filter-induced grouping) and local data patterns.

\textbf{Graph Representation}: The final step involves constructing the Mapper graph, consisting of nodes representing the clusters obtained previously:

$$
\{C_{i_1}, C_{i_2},\dots, C_{i_j} : 1\leq i \leq |I|\}.
$$

Edges between nodes \(C_{j_i}\) and \(C_{k_l}\) are established if their corresponding clusters share at least one common data point (i.e., \(C_{j_i} \cap C_{k_l} \neq \emptyset\)).

Mapper thus requires selecting a suitable filter function, clustering method, and input parameters \(\Gamma=(|I|, p, \theta)\). Here, \(|I|\) specifies the number of equally sized level sets, \(p\) determines the percentage of overlap between adjacent sets, and \(\theta\) represents clustering algorithm parameters. Notably, recent literature \citep{carr2021identifying} suggests that the choice of clustering algorithm itself minimally impacts Mapper's performance; instead, the pivotal factor is the carefully chosen filter function, influencing how data points cluster and connect.

Compared to traditional clustering techniques, Mapper has several distinct advantages. It incorporates domain-specific knowledge through the strategic selection of filter functions, guiding clustering toward meaningful structures. Moreover, Mapper avoids artificial partitioning of continuous data by using overlapping intervals, which preserves data continuity. Additionally, Mapper can systematically evaluate features by treating all variables as potential filters, aiding the identification of informative subpopulations.

\subsection{Persistence Diagrams via Super-level Set Filtration}

Persistent homology offers a structured approach to exploring the topological features of complex data spaces. Consider a compact set \(\mathcal{Z} \subseteq \mathbb{R}^D\). A filtration of \(\mathcal{Z}\) is an ascending sequence of subsets:

$$
\mathcal{Z}_1 \subseteq \mathcal{Z}_2 \subseteq \dots \subseteq \mathcal{Z}_n = \mathcal{Z},
$$

which incrementally reveals the space’s topology. Topological features—such as connected components, loops, and voids—emerge and disappear as the filtration progresses from \(\mathcal{Z}_i\) to \(\mathcal{Z}_j\).

Persistent homology quantitatively tracks these features' birth and death through persistence diagrams. A \(k\)-dimensional persistence diagram \(D_k\) records points \((b, d)\) in the birth-death plane, indicating a \(k\)-dimensional feature that appears at \(\mathcal{Z}_b\) and vanishes at \(\mathcal{Z}_d\). The persistence (lifespan) of each feature is defined as \(|d-b|\).

One widely utilized filtration approach is based on superlevel sets, defined with respect to a smooth function \(g: \mathcal{Z} \rightarrow \mathbb{R}\). For a value \(\alpha\in\mathbb{R}\), the superlevel set is given by:

$$
\mathcal{Z}_\alpha = \{z \in \mathcal{Z} : g(z) \geq \alpha\} = g^{-1}([\alpha, \infty]).
$$

As \(\alpha\) decreases from infinity to negative infinity, this defines an inclusion sequence \(\mathcal{Z}_{\alpha_1}\subseteq \mathcal{Z}_{\alpha_2}\), capturing evolving topological features. Critical values of \(g\) signal topological changes such as component merges or loop formations.

In our analysis, we utilize a Gaussian kernel density estimator (KDE) to construct the filtration function. Given a sample \(\hat{Z}_n=\{z_i\}_{i=1}^n\), the KDE is expressed as:

$$
g_n(y)=\frac{1}{n(\sqrt{2\pi}\eta)^D}\sum_{i=1}^{n}\exp\left(-\frac{\|y - z_i\|^2}{2\eta^2}\right), \quad y\in\mathbb{R}^D,
$$

where \(\eta>0\) is the kernel bandwidth. This KDE-based filtration effectively captures subtle density-driven topological features within the dataset.

\section{Proposed Methodology}
In this section, we propose an advanced methodological framework based on the Mapper algorithm, enriched by incorporating expert-driven insights. The key contribution of our approach lies in its ability to integrate domain-specific expertise into the Mapper analysis, significantly improving the interpretability and accuracy of results across various application areas.

The Mapper algorithm simplifies the analysis of complex, high-dimensional data by projecting it into a lower-dimensional representation using carefully selected filter functions. In our proposed methodology, these filter functions are explicitly designed in collaboration with domain experts, enabling the targeted highlighting of critical attributes relevant to the specific analytical context. For instance, these attributes could represent epidemiological indicators of disease spread, centrality metrics within social networks, or characteristic patterns of trading activities in financial markets. As a clustering method, Mapper effectively captures both local structures—by grouping data points exhibiting similar attributes—and global patterns via the filter functions, that might not be readily apparent through traditional clustering techniques, thereby providing greater interpretability than traditional clustering techniques.

\textbf{The Mapper-based Approach:} Our approach systematically identifies a subpopulation characterized by key attributes through an expert-informed Mapper analysis. The detailed procedural steps are outlined in Algorithm \ref{alg:mapper_intersection}:

\begin{algorithm}[H]
\caption{Identification of Prominent Subpopulations via Mapper and expert knowledge}
\label{alg:mapper_intersection}
\begin{enumerate}

    \item \textbf{Filter Function Definition:} 
    Define a set of \( n \) filter functions \( f_i \), each mapping agents to values corresponding to distinct symptom attributes, using expert knowledge. Specifically, the \( i \)th filter function \( f_i \) assigns each agent the value of the \( i \)th attribute.

    \item \textbf{Mapper Graph Construction:}
    Construct \( n \) Mapper graphs \( M_i \), each utilizing the corresponding filter function \( f_i \).

    \item \textbf{Target Level Set Identification:}
    For each Mapper graph \( M_i \), determine the target level set \( k \) exhibiting prominent symptom attribute characteristics.

    \item \textbf{Cluster Selection and Refinement:}
    For each Mapper graph \( M_i \), execute the following steps:
    \begin{enumerate}
        \item Identify clusters \( C_{k_1}, C_{k_2}, \dots, C_{k_j} \) associated with the chosen target level set \( k \).
        \item Refine clusters by removing soft clusters (clusters intersecting with clusters from other level sets). Formally, remove any cluster \( C_{k_m} \) if there exists a cluster \( C_{q_l} \) from another level set \( q \neq k \) satisfying \( C_{k_m} \cap C_{q_l} \neq \emptyset \).
        \item Let the agents within the remaining hard clusters form the subpopulation set \( S_i \).
    \end{enumerate}

    \item \textbf{Intersection of Subpopulations:}
    Derive the final subpopulation by computing the intersection across all refined subpopulations:
    \[
        Q = \bigcap_{i=1}^{n} S_i
    \]

\end{enumerate}
\end{algorithm}

A significant advantage of our proposed approach is its inherent iterative flexibility. Researchers can repeatedly apply the methodology to refine the analysis progressively by integrating additional expert insights or new datasets. For instance, a subpopulation identified through an initial analysis can subsequently serve as the input dataset for further iterations of the same methodological framework, enabling increasingly granular and high-resolution exploration without altering the fundamental methodological structure.

\section{Application to the Use of Insider Information in Stock Markets}

In this section, we apply the methodology described in Algorithm~\ref{alg:mapper_intersection} to identify insider trading in financial markets. Specifically, we utilize the previously defined framework to detect agents exhibiting trading behaviors symptomatic of informational advantages.

In financial markets, insiders—individuals with privileged access to material, non-public corporate information—may exploit their informational advantage by trading ahead of critical corporate announcements. Such insider-driven trades not only compromise market integrity but also yield unfair profits at the expense of uninformed market participants. We assume that company insiders are aware of forthcoming announcements, and these informed agents may disseminate first-hand information through social networks, allowing it to propagate further as second-hand information. According to Ref. \citep{ahern2017information}, inside tips typically originate from corporate executives and reach influential buy-side investors after approximately three links within the network.

Although we have no direct observations of individual agent types within these networks, the trading events themselves serve as observable symptoms indicative of informed behavior. We refer to the agents who trade using inside information as \emph{opportunistic agents}. To systematically identify these opportunistic agents who utilize insider information, we define two distinct filter functions aimed at capturing specific, suspicious trading patterns. Specifically, we consider two critical criteria: whether an agent consistently executes profitable trades just before public announcements, and whether the agent's trading performance significantly declines during other periods. 

Let $X_i$ represent the vector of realized returns for each transaction of the $i$-th investor in a given company. Moreover, $M_i$ denotes the total number of transactions, and we write $X_i = (R_{i1}, R_{i2}, \ldots, R_{iM_i})$, where $R_{ij} \in \mathbb{R}$ for all $j \in [1, M_i]$ denotes the realized return of the investor on transaction $j$. Each transaction $j$ of investor $i$ is classified based on whether it occurred during the pre-announcement period or outside of it. If transaction $j$ falls within the pre-announcement period, we denote $j \in A_i$; otherwise, $j \notin A_i$. The first filter function $F_1: X_i \rightarrow \mathbb{Z}$ is defined as
    \begin{equation}\label{eq:F1}
    F_1(X_i) = \sum_{j \in A_i} \mathbf{1}_{\{ R_{ij} > 0 \}}, \quad \text{where, } \mathbf{1}_{\{ R_{ij} > 0 \}} = 
    \begin{cases}
        1, & \text{if } R_{ij} > 0, \\
        0, & \text{otherwise}.
    \end{cases}
    \end{equation}
This filter counts the number of profitable transactions in the pre-announcement period for each investor. Although this filter function is necessary but not sufficient for identifying opportunistic agents. For instance, it gives a high score for agents who trade without any information and just happen to trade very often. We thus need to take into account agent-specific abnormal returns. Abnormal returns observed during the information event period, serve as the basis of $F_2$. The second filter function $F_2: X_i \rightarrow \mathbb{R}$ is defined as
    \begin{equation}\label{eq:F2}
    F_2(X_i) = \frac{1}{|A_i|} \sum_{j \in A_i} R_{ij} - \frac{1}{M_i - |A_i|} \sum_{j \notin A_i} R_{ij}.
    \end{equation}
This filter function measures the difference between the average return during pre-announcement periods and the average return during non-announcement periods for each investor. This criterion was identified, e.g., in \cite{givoly1985insider, fidrmuc2008insider, lakonishok2001insider}.  

Our hypothesis is that an opportunistic agent not only makes several profitable transactions before public announcements but also, their transactions during the pre-announcement periods are on average more profitable than otherwise.

\begin{remark}
    Additional filtering functions are possible. For example, using observations of the social graph of the agents, such as checking if the agent in question is connected to the insiders of a company they are trading in. 

\end{remark}
We chose to leave the social observations out of the clustering procedure. Instead, we use them to validate our method empirically. We do this by examining the hypothesis that compared to passive agents, the agents profiled as opportunistic, are more likely to be indirectly or directly socially connected to insiders and therefore more likely to receive insider information.

It is worth highlighting that our approach effectively addresses two key challenges:
\begin{enumerate}
    \item It enables the identification of groups of agents who exhibit suspicious trading behavior, especially during periods of new information releases.
    \item The within-bin clustering process inherent to our Mapper approach is particularly valuable, as it allows the detection of synchronized trading actions among agents within each cluster. This approach identifies groups of agents engaging in coordinated trading activities, whether profitable, unprofitable, or no-trade cases, along with the specific timing of these trades.
\end{enumerate}
This dual capability of detecting both suspicious behavior and synchronization in trading provides a robust framework for analyzing agent activities, making it a powerful tool in financial market surveillance.

\subsection{Motivation: Mapper vs. Traditional Clustering}

We now present a practical example demonstrating why the Mapper framework provides distinct advantages over traditional clustering techniques such as K-means. Consider three hypothetical agents $\{A_i~i \in \{1,2,3\}\}$ with distinct trading behaviors represented through their respective return profiles:
\[
A_1 = [0.005, 0, 0, 0, 0, 0, 0],\quad
A_2 = [0, 0, 0, 0, 0, 0.005, 0],\quad
A_3 = [0, -0.005, 0, 0, 0, 0, 0]
\]

First, we assess the similarity of these agents by calculating pairwise Euclidean distances between their return profiles:

\[
\text{ED}(A_1, A_2) = \sqrt{0.01323},\quad
\text{ED}(A_1, A_3) = \sqrt{0.01323},\quad
\text{ED}(A_2, A_3) = \sqrt{0.01}
\]


Traditional clustering methods such as K-means rely heavily on geometric distances like Euclidean distance. Given these computed distances, K-means clustering would group agents primarily based on numerical proximity, placing \( A_2 \) and \( A_3 \) together due to their smallest mutual distance. However, from a domain-specific perspective, such grouping might be counterintuitive. Although \( A_2 \) and \( A_3 \) are numerically closer, their trading behaviors differ significantly in meaningful ways: \( A_1 \) and \( A_2 \) both exhibit distinct positive returns, suggesting a similar underlying behavioral trait (potential informed trading), while \( A_3 \) displays predominantly neutral or negative behavior indicative of passive or uninformed trading.

This reveals a critical limitation of direct clustering methods, such as K-means, which rely purely on geometric proximity without accounting for temporal or domain-specific behavioral context. The Mapper algorithm overcomes these limitations by incorporating a topological approach that analyzes the shape of the data in a multi-dimensional feature space. Rather than clustering based solely on distance metrics, Mapper first applies a filter function to capture key aspects of the data, such as return, volatility, trading frequency, or other domain-specific metrics. This filter function transforms the data into a simplified representation that emphasizes behavioral patterns rather than raw numerical distances.
In the example, the Mapper algorithm can differentiate \( A_1 \) and \( A_2 \) as engaging in informed trading while isolating \( A_3 \) as uninformed, addressing the core limitation of K-means in this context.

\section{Results}

\subsection{Monte-Carlo Experiment}

To validate our method, we conducted simulations on synthetic data where the true status of each agent was known. A cascade model was simulated on an actual empirical insider graph along with the baseline model where no information prevails in the market.  
We compared the Mapper approach with several alternative clustering techniques including K-means, K-means++, DBSCAN, Hierarchical Clustering (Hclust), and anomaly detection methods including One-Class Support Vector Machines (OCSVM), K-Nearest Neighbors (KNN), Isolation Forest (IForest), and Local Outlier Factor (LOF). 
Our approach demonstrated superior performance, particularly in minimizing false positives. A conservative approach is needed in applications such as insider trading detection, where false accusations have significant consequences. 
Our method maintained a high precision across different baseline distributions, whereas other methods struggled with increased false positives as similarity between distributions of behaviors increased. The results are robust with different settings. The data generation settings, the benchmark models, and results from Monte-Carlo Experiment are reported in \ref{app_monte}.

\subsection{Empirical Data}

In our empirical analysis, we assume that information spreads before the arrival of a public company announcement, resulting in a cascade of opportunistic trading. The hypothesis is that information may begin to spread from a company's board through social links prior to an announcement. To analyze this hypothesis, we use a unique and rich data set combined from multiple sources: (i) Data that contains the insider network in Finland: Being an insider in the same company creates a social link between all members. Overlapping board memberships thus result in a large social network; (ii) Information of board members and insiders' mandatory disclosure notifications, which allowed us to identify partial trading patterns; (iii) A unique and extensive pseudonymized dataset containing detailed trading records of all investors across all securities on the Helsinki Stock Exchange; (iv) Data related to company announcements. By integrating these datasets, we have created unique data that contains not only the social links between insiders but also track all their transactions across all securities extending beyond those restricted by insider trading regulations (see our recent paper \cite{baltakys2023predicting}). To ensure compliance with relevant legislation, all personal data within the dataset is pseudonymized, we do not access any information that identifies the individuals. 

Descriptive statistics for the data are presented in Table \ref{descriptive_statistics} (See \ref{appdata} for details on data preprocessing). The data encompasses the period from 05/2005 to 12/2009, which includes 1,159 trading days across 122 companies. We designated each trading day as occurring either inside or outside a pre-announcement period for a given company, based 
on whether an unscheduled announcement was made by the company within four days. 
There were 1,596 identified investors. We removed trades in companies that do not have any trades during pre-announcement periods. The remaining data consists of 119 companies and 1586 investors, and 1490 of the investors made at least one trade during pre-announcement periods and one outside these periods, and we labeled these as active. There were a total of 293,750 transactions, with 81,634 occurring within the pre-announcement periods. An average investor was socially connected to 17.3 other investors in the insider network. Approximately 22\% of the investors were female. 

There is a significant variation in the liquidity of companies: the most liquid companies are traded nearly every day by at least one investor included in our analysis, while the least liquid are traded only on a few days per year. This is mirrored in the transaction numbers for different companies, with the highest being 3,544 and the lowest around 2. On average, each company made 19 announcements per year. However, the frequency of announcements varied significantly, with some companies having as few as 2 announcements per year, while others had as many as 69 announcements annually. For an average investor, around 23\% of transactions were located within the pre-announcement periods. 24.49\% of their profitable and 22.02\% unprofitable transactions were located within the pre-announcement periods. Out of an average of 669 transactions, 339 were profitable, which is very close to 50\%. The table also shows the age distribution. 

Due to variations in the number of announcements across different companies, the number of pre-announcement days also varies for each company. We find that a majority of the companies were quite illiquid in terms of the number of transactions, and for most companies, the pre-announcement days constituted about 20-40\% of all trading days. To gauge the robustness of our analysis, we conduct additional assessments by filtering companies based on transaction volume. 

\begin{table}[ht!]
  \centering
    \begin{adjustbox}{max width=\textwidth}
    \begin{tabular}{lccccccc}
    \toprule
    \textbf{Descriptive Statistics} &&&&&& \\
    \midrule
    Number of trading days & \multicolumn{2}{c}{1,159}&&&& \\
    Number of trading days per year & \multicolumn{2}{c}{251 }&&&&\\    
    Number of companies & \multicolumn{2}{c}{122} &&&&\\
    Number of investors & \multicolumn{2}{c}{1,596}&&&& \\
    Total number of transactions &\multicolumn{2}{c}{ 293,750}&&&& \\
    Total transactions within pre-announcement periods & \multicolumn{2}{c}{81,634} &&&&\\
    Average Degree & \multicolumn{2}{c}{17.3}&&&& \\
    Fraction of Female Investors  & \multicolumn{2}{c}{21.65\%}&&&& \\

    \midrule
         Company specific statistics & \multicolumn{1}{c}{Min.} & \multicolumn{1}{c}{1st Qu.} & \multicolumn{1}{c}{Median} & \multicolumn{1}{c}{Mean} & \multicolumn{1}{c}{3rd Qu.} & \multicolumn{1}{c}{Max.} & \multicolumn{1}{c}{Std Dev} \\
          \midrule
    Number of trading days with transactions/company/year & 0.87  & 30.13 & 56.69 & 72.38 & 105.69 & 226.53 & 56.12 \\
    Number of transactions/company/year & 2.38  & 88.33 & 209.22 & 520.95 & 707.07 & 3544.63 & 686.66 \\
    

    Number of announcements/company/year & 1.95& 12.82 & 17.53& 19.24 & 23.48& 69.45 & 9.89    \\
    \midrule
    Investor specific statistics& \multicolumn{1}{c}{Min.} & \multicolumn{1}{c}{1st Qu.} & \multicolumn{1}{c}{Median} & \multicolumn{1}{c}{Mean} & \multicolumn{1}{c}{3rd Qu.} & \multicolumn{1}{c}{Max.} & \multicolumn{1}{c}{Std Dev} \\
          \midrule
  Investor's unique trading days&  1 &  3 &  10 &  30.51&  27.25& 654 & 60.64\\
  Number of investors traded/company/year &  0.6 &  12 &  20.10 &  44.27 &  56.85 & 268.2 &  51.79  \\
  Investor's all transactions/company& 11.00 & 408.25 & 967.00 & 2407.79 & 3268.00 & 16383.00 & 3173.69\\
Investor's all transactions/company  & 0.00 & 69.50 & 192.00 & 669.13 & 936.00 & 3839.00 & 948.89\\
\quad within pre-announcement periods &  &  &  &  &   &  &  \\
Percentage of transactions within  & 0.00\% & 15.57\% & 22.14\% & 23.14\% & 29.32\% & 60.53\% & 11.24\%\\
\quad pre-announcement periods per company\\

 Investor's all profitable transactions per company&  1.00 & 163.25 & 452.50 & 1188.47 & 1808.50 & 8517.00 & 1590.88\\ 
Investor's profitable transactions per company& 0.00 & 33.00 & 92.50 & 339.33 & 479.25 & 2271.00 & 490.09 \\
\quad within pre-announcement periods &  &  &  &  &   &  &  \\
Percentage of profitable transactions & 0.00\% & 15.82\% & 23.97\% & 24.49\% & 31.56\% & 68.75\% & 12.75\%\\
\quad within pre-announcement periods per company\\
    \midrule
    Age Group &       < 30 & 31-40&   41-50 & 51-60 & > 61 &&\\
    \midrule
  & 3.49\% & 9.00\% & 32.02\% & 26.63\% & 15.35\% & &\\
    \bottomrule
     \end{tabular}%
     \end{adjustbox}
     \caption{ \label{descriptive_statistics}Descriptive statistics of the empirical data.}
     
\end{table}%
\subsection{Empirical findings on insider network}

\begin{figure}[!ht]
\begin{center}
    \subfigure[Kernel Density Plots.]{%
        \includegraphics[scale=.30]{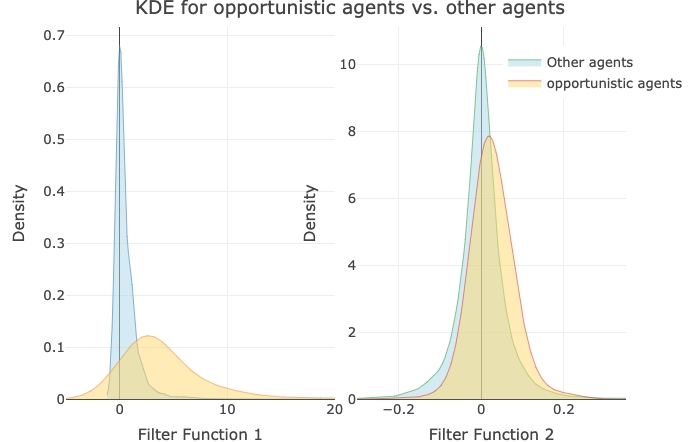}
        \label{fig:kde}}
    \quad
	\subfigure[Network Visualization. ]{%
        \includegraphics[scale=.22]{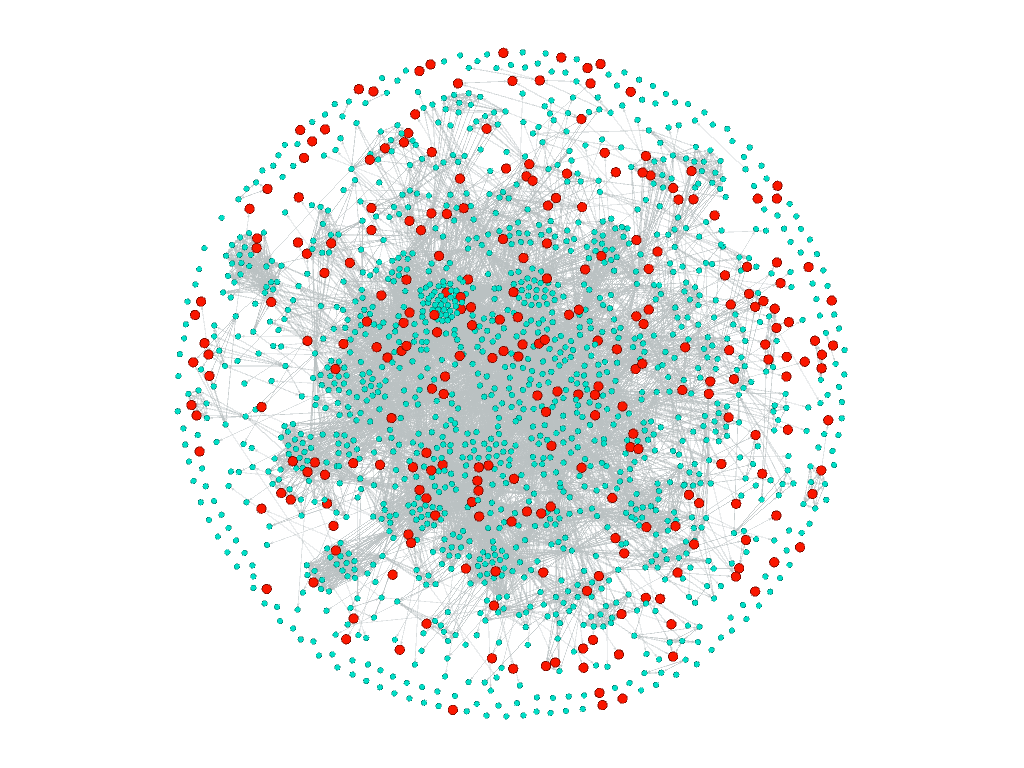}
        \label{fig:network}}    
    \quad
        \subfigure[Histograms]{%
        \includegraphics[scale=.7]{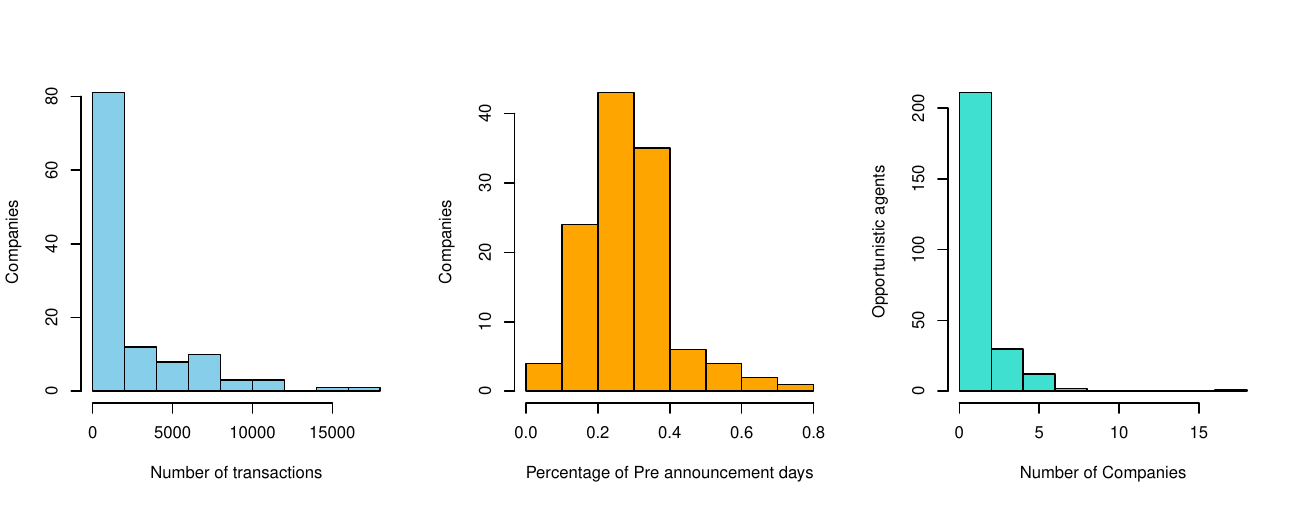}
        \label{fig:histogram}}
\end{center}
\caption{
\textbf{(a)} 
\textbf{Kernel Density Plots}: Comparison of Filter function densities for opportunistic agents and other agents. $F_1$ represents the number of profitable trades during the pre-announcement period, while $F_2$ measures the difference in realised returns between the announcement and non-announcement periods. The plots provide a visual overview of the distributions, offering insights into the trading behavior of both opportunistic agents and other agents in response to information received. 
\textbf{(b)} 
\textbf{Network visualization}: nodes are the agents with edges as links between them. The opportunistic agents as identified by the Mapper approach are highlighted in red and the others are turquoise colored.}\label{fig:results_combined}
\textbf{(c)} 
\textbf{The set of histogram plots } Read from left to right: 
(1) Number of transactions per company. (2) Percentage of pre-days of companies with respect to the total number of transactions executed by all the investors. 
(2) The third plot depicts the distribution of companies with respect to the percentage of 
days within the pre-announcement periods out of all the trading days. 
(3) The histogram illustrates the distribution of opportunistic agents, showing the number who traded one company opportunistically, those who traded two companies opportunistically, and so on. 
\end{figure}

The kernel densities for both opportunistic agents and other agents as identified using the filter functions $F_1$ and $F_2$ (Eqs. \ref{eq:F1} and \ref{eq:F2}) are presented in Figure \ref{fig:kde}. The pattern is clear: during the pre-announcement period, opportunistic agents show a large number of profitable trades. These agents also exhibit higher returns in the pre-announcement period compared to non-announcement periods, as indicated by the right-skewed plot of filter function 2. For other agents, the values of $F_1$ and $F_2$ cluster around zero. 

The KDE curves confirm that the agents identified as opportunistic are not simply extreme or outlier values in either filter function. The distributions for these agents are broader and more varied, rather than concentrated at the high ends of the scale. The wider spread in both $F_1$ and $F_2$ indicates that these agents are characterized by a consistent pattern of profitability and abnormal returns during pre-announcement periods, but not exclusively due to having the highest values. The method captures a range of behaviors associated with opportunism, rather than just picking up outliers. This suggests that the Mapper method identifies agents whose overall patterns across both profitability and timing of transactions indicate informed and deliberate trading actions.

The rightmost plot in Figure \ref{fig:histogram} shows the distribution of agents identified as opportunistic in terms of how many companies they traded opportunistically on. Only a handful of agents behaved opportunistically with respect to more than just a few companies. Therefore, the opportunistic behavior appears to be very company-specific. This is reasonable because presumably, very few investors have access to insider information on multiple companies. 

A visualization of the insider network, highlighting agents identified as opportunistic in red, is shown in Figure \ref{fig:network}. The opportunistic agents are primarily located in the giant component, but scattered around the network. This suggests that agents who share private information without taking advantage of it themselves may play a crucial role in spreading information across the network. This aligns with the findings of \cite{ahern2017information}, which show that buy-side traders account for the majority of tippees only after an average of three links in the network. This suggests that private information is not exploited in the immediate vicinity of its original source. It is important to keep in mind that agents’ behavior may well reflect the presence of social connections not captured by our data---for instance, connections based on friendships or memberships in organizations such as golf clubs.

The empirical findings of the Mapper method are summarised in Table \ref{tab:results_mapper}, including key statistics including profits per agent during both the announcement and non-announcement periods for opportunistic agents. Their transactions’ profitability is compared with that of the remaining agents. Altogether, 1,586 investors, 119 companies, and 15,668 investor-company pairs (a pair occurs if a specific investor trades a specific stock) were present in our analysis. The distribution of companies with respect to the total number of transactions executed by all investors and with respect to the percentage of pre-announcement trades is shown in the first two plots from the left-hand side in Figure \ref{fig:histogram}. It is noteworthy that, in the case of all companies, the least liquid company had 59 transactions in total. To control for the effect of liquidity, we filtered the companies based on the number of total transactions, using thresholds of 5,000, 6,000, and 7,000 transactions. This resulted in 24, 18, and 11 companies; 1,215, 1,179, and 1,112 investors; and 8,311, 7,169, and 3,532 investor-company pairs, respectively. 

In the data for all 119 companies and 1,586 investors as agents in the network, a subset of 256 (16.14\%) agents were identified as opportunistic. About half of the agents identified as opportunistic were profiled based on trading with less liquid stocks. When we applied the method to the most liquid companies, the proportion of opportunistic agents decreased---when the analysis is done with the 11 most traded companies that represent also the most liquid stocks, only 4.23\% of investors are identified as opportunistic ones. This suggests that the opportunistically behaving investors are related to the trading of less-liquid and smaller companies. This result aligns with Ref. \citep{lakonishok2001insider}, which states that insiders are better at predicting aggregate movements of small companies than those of large companies. This, in turn, incentivizes leveraging insider benefits specifically with smaller companies. 

When analyzing whether agents are connected to the companies they trade in through the insider network, we found that, using all the data, 342 out of 454 identified opportunistic agent-company pairs (75.3\%) were interconnected with a maximum seed distance of 6. Similarly, among the 15,214 non-opportunistic agent-company pairs, 11,090 pairs (72.9\%) were interconnected with a maximum seed distance of 8. This association becomes more evident when less-liquid companies are filtered out. After filtering, \textit{all opportunistic agent-company pairs} (100\%) are connected to the seeds, while for non-opportunistic agents, approximately 77\% of pairs remain interconnected. We consider this a strong result. Despite not using information about the insider network in clustering, we observed a significant pattern: opportunistic agents are far better connected to company insiders compared to other agents. This strongly suggests that the identified agents are genuinely opportunistic. The finding is stronger for liquid companies, as they tend to be more central within the insider network. To account for this effect, we analyzed the proportion of investors connected to the companies they trade with within no more than four steps, and the results remained consistent. Although opportunistic agents maintain full connectivity in the most liquid companies, their influence in terms of euro volume during pre-announcement periods diminishes sharply from 65\% in the top 24 companies to 25\% in the top 11 most traded companies. This suggests that heightened regulatory scrutiny and market surveillance in these highly liquid companies reduce the ability of opportunistic agents to exploit informational advantages. 

Given that the first filter function measures the number of profitable trades made in the pre-announcement periods, and that the second filter function measures the difference between their average daily returns in the pre-announcement periods, we analyze the results in terms of related, but not exactly similar measures. Firstly, we observe that compared to other agents, agents identified as opportunistic concentrate their transactions, in terms of euro volume, on pre-announcement periods. This is a particularly strong result when data on all companies is used, and the difference also persists when less liquid companies are filtered out. Secondly, we observe that for opportunistic agents, 38\% of profitable transactions and 24\% of unprofitable transactions were executed within the pre-announcement periods, while the percentage is a flat 28\% for the other agents. This is a robust result with respect to different data filtering thresholds. Thirdly, agents identified as opportunistic generate significantly higher euro profits during pre-announcement periods, especially in the most liquid companies (e.g., 22,322 euros in the top 24 companies). By contrast, opportunistic agents generate negative or minimal profits during non-announcement periods, especially in the most liquid companies (e.g., -3,005.4 euros in the top 24 companies). This suggests that profitability is highly concentrated around informational events, suggesting a reliance on pre-announcement trading strategies rather than sustained performance across all periods. The significant disparity in average euro profit between pre- and non-announcement periods (e.g., 25,327.4 euros in the top 24 companies) is suggestive of opportunistic agent's reliance on exploiting information asymmetry. This indicates that we have identified agents who are not merely participating in the market but are systematically exploiting specific periods \footnote{The details on the attributes shared by the opportunistic investors are presented in \ref{app_add}}. 

In Ref. \citep{baltakys2023predicting}, investors’ trading decisions for the next day were predicted from how their social neighbors traded today, using Graph Neural Networks. The goal of that paper was to identify investors whose trading decisions are influenced by their social connections. Investors whose decisions were highly predictable, were considered to be behaving suspiciously. They used almost the same data, and an interesting question is whether these two completely different approaches yield overlapping results. Our current research shares 27 of these companies, for which there were 1,284 investors in total. Out of them, we identify 125 agents as opportunistic, while 158 were identified as suspicious in Ref. \citep{baltakys2023predicting}, with an overlap of 50 investors. Using Fisher’s Exact Test (based on the hypergeometric cumulative distribution), the p-value for the overlap is less than $10^{-10}$, indicating an extremely strong statistical significance. This means that such overlap cannot be explained by randomness. This shows a considerable alignment between the two completely different approaches, which further verifies both the methods presented in this paper and in Ref \citep{baltakys2023predicting}.

We additionally ran DBSCAN clustering on the empirical data, as it yielded the second-best results in the simulated data scenario, following the Mapper approach. DBSCAN identified 951 agents as opportunistic out of a total of 1,284, with an overlap of 112 traders with the 125 opportunistic agents identified by the Mapper algorithm and an overlap of 130 agents with the 158 suspiciously behaving agents identified in \cite{baltakys2023predicting}. Notably, the overlap between DBSCAN and other methods is higher not due to superior accuracy, but rather because DBSCAN tends to cluster a larger number of agents as opportunistic agents, indicating a higher error rate. This divergence suggests that DBSCAN’s clustering results are less precise compared to the Mapper algorithm, as the method appears to be overestimating the number of opportunistic agents. This aligns strongly with the observation made with the simulated data.


\begin{table}[ht!]
    \begin{adjustbox}{max width=\textwidth}

  \centering
 
  \begin{tabular}{lrrrrrrrr}
    \toprule
    \textbf{ } & \multicolumn{2}{c}{\textbf{All data}} & \multicolumn{2}{c}{\textbf{Data for 24 most}} & \multicolumn{2}{c}{\textbf{Data for 18 most}} & \multicolumn{2}{c}{\textbf{Data for 11 most}} \\
    \textbf{ } & \multicolumn{2}{c}{\textbf{}} & \multicolumn{2}{c}{\textbf{ traded companies}} & \multicolumn{2}{c}{\textbf{ traded companies}} & \multicolumn{2}{c}{\textbf{ traded companies}} \\
    
    \midrule
    \textbf{Companies} & \multicolumn{2}{c}{119} & \multicolumn{2}{c}{24} & \multicolumn{2}{c}{18} & \multicolumn{2}{c}{11} \\
    \textbf{Number of investors} & \multicolumn{2}{c}{1,586} & \multicolumn{2}{c}{1,217} & \multicolumn{2}{c}{1,179} & \multicolumn{2}{c}{1,112} \\
    \textbf{Number of investor-company pairs} & \multicolumn{2}{c}{15,668} & \multicolumn{2}{c}{8,311} & \multicolumn{2}{c}{7,169} & \multicolumn{2}{c}{4,532} \\
    \textbf{Minimum number of transactions} & \multicolumn{2}{c}{59} & \multicolumn{2}{c}{5,000} & \multicolumn{2}{c}{6,000} & \multicolumn{2}{c}{7,000} \\
    
    \midrule
   \textbf{
   } &\multicolumn{1}{c}{\textbf{Opportunistic}} & \multicolumn{1}{c}{\textbf{Others}} & \multicolumn{1}{c}{\textbf{Opportunistic}} & \multicolumn{1}{c}{\textbf{Others}} & \multicolumn{1}{c}{\textbf{Opportunistic}} & \multicolumn{1}{c}{\textbf{Others}} & \multicolumn{1}{c}{\textbf{Opportunistic}} & \multicolumn{1}{c}{\textbf{Others}} \\
   \midrule
    \textbf{Investors}  & 256   & 1,330  & 126   & 1,091  & 123   & 1,056  & 47    & 1,065 \\
    \textbf{}  & (16.14\%)	& (83.86\%)	& (10.35\%)	& (89.65\%)	& (10.43\%)	& (89.57\%)	& (4.23\%)	& (95.77\%) \\
   \multicolumn{1}{p{5cm}}{ \textbf{Percentage of connected investor-company pairs}
     } & 75.3\%  & 72.9\%  & 100\%  & 77.6\%  & 100\%  & 77\%  & 100\%  & 77\% \\

  \multicolumn{1}{p{5cm}}{ \textbf{Percentage of connected investor-company pairs within 4 steps}
     } & 71.4\%  & 70.6\%  & 77.66\%  & 74.7\%  & 78.4\%  & 75.16\%  & 81.8\%  & 75.16\% \\

   \midrule       
   \multicolumn{1}{p{5cm}} {   \textbf{Fraction of euro volume in pre-announcement periods} }& 57\%  & 51\%  & 65\%  & 60\%  & 27\%  & 26\%  & 25\%  & 23\% \\
   \multicolumn{1}{p{5cm}} {\textbf{Fraction of profitable transactions in pre-announcement periods vs all profitable transactions}} & 38\%  & 28\%  & 39\%  & 31\%  & 38\%  & 31\%  & 33\%  & 29\% \\
    \multicolumn{1}{p{5cm}}{ \textbf{Fraction of unprofitable transactions in pre-announcement periods vs all unprofitable transactions} }& 24\%  & 28\%  & 26\%  & 31\%  & 26\%  & 31\%  & 21\%  & 30\% \\
     \multicolumn{1}{p{5cm}} { \textbf{Euro profit in pre-announcement periods per investor} }& 11,991.5 & -2,461.0 & 22,322.0 & -1,455.6 & 933.6 & 92.0  & 920.8 & 164.8 \\
   \multicolumn{1}{p{5cm}} {   \textbf{Euro profit in non-announcement periods per investor} }& 94.8  & 3,079.7 & -3,005.4 & 3,963.6 & -1,072.3 & 322.1 & -2,847.7 & 340.4 \\

  \multicolumn{1}{p{5cm}} {   \textbf{Difference of Average Euro profit pre and non-announcement periods per investor}} & 11,896.7 & -5,540.7 & 25,327.4 & -5,419.2 & 2,005.9 & -230.1 & 3,768.5 & -175.6 \\ \hline

    \bottomrule

    \end{tabular}%
    \end{adjustbox}

     \caption{Performance Metrics Comparison between opportunistic and other agents. The table presents results for different cases of analysis. The first column contains the whole data, and from left to right, we filter out companies with fewer than 5,000, 6,000, and 7,000 transactions respectively. The percentage of connected investor-company pairs reports the proportion of pairs where an investor is directly or indirectly connected to a company they traded in, in the insider network.}

  \label{tab:results_mapper}%
\end{table}%

\subsection{Statistical validation using the topological features of the data}


The sparse nature of our data makes statistical validation challenging for conventional distribution models. 
To address this issue, we employ TDA to establish the statistical validity of our findings by examining the topological characteristics of the data. Our aim is to examine, whether the topological properties of agent transactions differ between pre-announcement and non-announcement periods for the agents that were identified as opportunistic by Mapper, as compared to other agents. The intuition is that opportunistic agents are more likely to alter their trading behavior when new information is about to be released, due to their access to inside information through their social connections.

TDA involves the study of point cloud data, which consists of a set of data points in a given coordinate system with a defined distance or similarity measure. By examining the structural properties of this data, such as connectivity and shape, TDA provides a way to analyze the geometric and topological characteristics of the data beyond mere numerical values. One key tool in TDA is homology, an algebraic method for examining the topological features of a data space. Homology identifies $k$-dimensional holes, specifically $k=0$ corresponds to connected components, $k=1$ to loops, and $k=2$ to voids. Persistent homology builds on this by illustrating how these topological features evolve over different scales. Specifically, we apply this method to analyze the super-level sets of real-valued functions $g$ defined over a space $\mathz$, denoted as $\mathz_\alpha = \{z \in \mathz : g(z) \geq \alpha\}$.

The $k$-th homology group $H_k$, for $k = 0, \ldots, \text{dim}(\mathz)$, describes the $k$-dimensional holes, and its rank, known as the Betti number $\beta_k$, counts the number of these holes. The Betti numbers often serve as a summary of the topology of each $\mathz_\alpha$. Specifically, $\beta_0$ represents the number of connected components in $\mathz_\alpha$, while, in general, $\beta_k$ counts the number of $k$-dimensional holes in $\mathz_\alpha$. Persistent homology improves this approach by tracking how these homological features persist and change as the parameter $\alpha$ varies. This method provides a more dynamic view of the topology, allowing us to observe the birth and death of features across different scales. This gives insights into the multi-scale structure of the data that basic homology alone cannot provide. By taking into account the persistence of topological features, persistent homology captures not just a space's static topological features but also how these properties evolve, providing a more complete knowledge of the underlying form and structure.

Persistent homology is unquestionably the most popular method in the rapidly expanding area of TDA, thanks in large part to its ease of visualization via persistent barcodes and diagrams. In the context of super-level set filtration, each bar in a persistence barcode represents an interval that starts (is `born') at a level $\alpha = b$, where a new feature in the homology of $\mathz_\alpha$ emerges, and ends (dies) at a lower level $\alpha = d < b$, where this feature disappears \footnote{In this sequel, we consider analyzing the persistent homology for zero-dimensional and one-dimensional features i.e. we take $k=0,1$.}. Persistence diagrams, which plot the points $(b, d)$, give an analogous depiction of barcodes. 

The parametric model for a persistence diagram can be obtained by using Gibbs distribution. We refer to this approach as Gibbs distribution for persistence diagrams (GPD). The subsequent description outlines a procedure for a set of $N$ points on a persistence diagram $D_k$ for fixed homology group $k$.

 Given the set $\mathcal{Z}_N=\{b_i,d_i\}_{i=1}^N$ of birth death pairs on the persistence diagram $D_k$, construct a new set ${\mathcal{X}}_N = \{{x}_i\}_{i=1}^N$, with ${x}_i^{(1)} = d_i$ and ${x}_i^{(2)} = b_i - d_i$. That is, ${\mathcal{X}}_N$ is a set of $N$ points in $ \mathbb{R} \times \mathbb{R_+}$. This transformation projects the diagonal line on $D_k$ on the horizontal axis resulting in a projected persistence diagram $\text{PD}_k$. 
Let ${x}_q \in {\mathcal{X}}_N $, for $q \geq 1$ denotes the $q$-th nearest neighbor of ${x}$, for any ${x} \in {\mathcal{X}}_N  $. Define $\mathbb{L}_q(\tilde{x}_N)$ as
$$
\mathbb{L}_q({\mathcal{X}}_N) =\sum_{{x} \in {\mathcal{X}}_N} \| {x}-{x}_q\|.
$$

The statistical model for $\text{PD}_k$ is represented by the Gibbs distribution, given as follows:

$$
\phi_\Theta(\mathcal{X}_N) = \frac{1}{Z_\Theta} \exp\left(-H_\Theta^K(\mathcal{X}_N)\right),
$$

where $\Theta = (\theta_0,\theta_1, \ldots, \theta_K)$ and $H_\Theta^K(\mathcal{X}_N):\mathbb{R}^N \rightarrow \mathbb{R}$ is defined for $K \geq 1$ as:
$$
\tilde{H}_\Theta^K(\mathcal{X}_N) = \sum_{q=1}^{K} \theta_q \mathbb{L}_q(\mathcal{X}_N) \times (g_n(\mathcal{X}_N))^{\theta_0}.
$$

Since the analytical form of $Z_{\Theta}$ is not known, the estimation of parameters is done using the pseudo-likelihood approach \citep{adler2017modeling}.
The pseudo-likelihood function is given by
$$
\mathcal{L}^K_{\Theta}({\mathcal{X}}_N) = \prod_{x \in {\mathcal{X}}_N} f_{\Theta}(x|{N}_K(x))
$$
where $N_K(x)$ denotes the $K$ nearest neighbors of $x$ in ${\mathcal{X}}_N$, and
$$
f_{\Theta}(x | N_K(x)) = \frac{ \exp\left(-\tilde{H}_\Theta^K(x | {N}_K(x))\right)}{\int_{\mathbb{R}} \int_{\mathbb{R_+}} \exp\left(- \tilde{H}_\Theta^K(z | {N}_K(x))\right) \, dz^{(1)} \, dz^{(2)}}
$$
 with 
$$
\tilde{H}_\Theta^K(x | {N}_K(x)) = \sum_{q=1}^{K} \theta_k \mathbb{L}_k(N_K(x))\times ({g}_n(x)^{\theta_0}
$$
The parameter $\theta_0$ is a non-negative parameter. Note that nearest neighbors capture the closeness relations between the points, and ${g}_n$  controls the shape of the whole points on the persistence diagram. The parameter set $\Theta$ is learned by minimizing the negative log-likelihood, which is given by
\begin{align}
-\log(\mathcal{L}^K_{\Theta}({\mathcal{X}}_N)) 
&= -\sum_{x \in {\mathcal{X}}_N} \log\left( f_{\Theta}(x|N_K(x)) \right) \notag \\
&= -\sum_{x \in {\mathcal{X}}_N} 
\log\left( \exp\left(- \tilde{H}_\Theta^K(x | N_K(x))\right) \right) \notag \\
&\quad +  \log\left( \int_{\mathbb{R}} \int_{\mathbb{R_+}}  
\exp\left(-\tilde{H}_\Theta^K(z | N_K(x))\right) \, dz^{(1)} \, dz^{(2)} \right) 
\end{align}
So, the simplified negative log-likelihood is

\begin{equation}
\sum_{x \in \tilde{x}_N} \tilde{H}_\Theta^K(x | N_K(x)) + \sum_{x \in \tilde{x}_N} \log\left( \int_{\mathbb{R}} \int_{\mathbb{R_+}}  \exp\left(- \tilde{H}_\Theta^K(z | N_K(x))\right) \, dz^{(1)} \, dz^{(2)} \right) \notag
\end{equation}


This parametric model allows to model $\text{PD}_k$ obtained using a given data sample. The primary motivation for using GPD for statistical validation of our findings is attributed to our constrained data scenario, wherein we possess only a single sample of the investor data. GPD is specifically tailored to function effectively with a single persistence diagram, making it a suitable choice for our analysis. In order to do so, we define distributions over the space of persistence diagrams for opportunistic agents during both pre-announcement and non-announcement periods. The aim is to observe statistical differences in these distributions and to apply the same analysis to other types of agents as well. Based on the GOD model, we can check if two data sets come from the same distribution by fitting a model to the projected persistence diagrams of each data set and then comparing the estimates. We do this by testing the hypothesis $H_{null}: \theta_j^1 = \theta_j^2$ vs. the alternative hypothesis $H_{alt} : \theta_j^1 \neq \theta_j^2$, for $j = 0,1, \ldots, K$, where $\Theta_i=(\theta^i_1,\theta^i_2,\ldots,\theta^i_n)~ i\in \{1,2\}$ are parameters for the fitted models for two data sets. The test-statistics is given by
\begin{equation}\label{eq:test}
T=\left|\frac{\hat{\theta}_j^{1} - \hat{\theta}_j^{2}}{\sqrt{\text{Var}(\hat{\theta}_j^{1}) + \text{Var}(\hat{\theta}_j^{2})}}\right|
\end{equation}

\noindent
where $\hat{\theta}_j^{s}$ is the estimated $\theta_j^{s}$ for $s \in \{1,2\}$. Based on the asymptotic properties of the pseudo-likelihood, the p-value is obtained by
\[
2 \times P\left(Z > \left|\frac{\hat{\theta}_j^1 - \hat{\theta}_j^2}{\sqrt{\text{Var}(\hat{\theta}_j^1) + \text{Var}(\hat{\theta}_j^2)}}\right|\right), 
\]
where $Z$ is the standard normal random variable. The variances of the estimates are based on the empirical Fisher information matrix.
In the present paper, we have chosen to set a maximum value for $K=3$ as prior research has demonstrated that increasing the value of $K$ beyond 3 yields only marginal improvements \citep{adler2017modeling}.\footnote{The code for estimating the parameters of GPD is available on GitHub \url{https://github.com/Anubha0812/GPD}.}


\begin{remark} The GPD method alone cannot directly identify opportunistic agents. The statistical differences observed between the PDs during pre-announcement and non-announcement periods do not, by themselves, indicate opportunistic trading behavior. Instead, we apply the GPD method separately to investor categories identified by the Mapper algorithm. This approach allows us to analyze whether opportunistic agents exhibit topological properties in their behavior that differ from other agents. For instance, consider two investors: one exhibits high trading activity outside the announcement period, while the other intensifies trading before announcements. Both show statistically significant differences between their persistent diagrams in the pre-announcement and other periods. However, these differences alone do not distinguish between the opportunistic and non-opportunistic agent. The first investor, who trades heavily outside the announcement period, is not opportunistic, whereas the second, who strategically increases activity before announcements, is.  
Statistical significance in persistent diagrams must be interpreted within the context of investor behavior to identify true opportunism. 

Moreover, this method is far more computationally demanding. Given the vast amount of data, applying GPD broadly would require impractical computational resources, greatly impacting efficiency without offering substantial additional insights. By focusing on a pre-selected subset of agent-company pairs, our targeted approach balances statistical rigor with computational feasibility, making it both effective and efficient.
\end{remark}
The first step in analyzing the topology of transaction space is to obtain PDs using the point cloud of transaction data for the investors in both the pre-announcement period and the non-announcement period. To construct the point cloud, we use two key metrics: daily transaction realised returns and profits. The point cloud for the announcement period for an investor $i$ is represented by the matrix 
$ X_i = [R_{ij}, P_{ij}]_{j \in A_i}. $
Here,  $P_{ij}$ represents the euro profit from transaction $j$, where profit is computed by multiplying the realized return by the transaction volume (in Euros). PDs are constructed on the top of these point clouds using super-level set filtration. To illustrate, we present the PDs of randomly selected opportunistic agents during both the announcement and non-announcement periods with barcodes that depict the persistence of the topological features in \ref{app_add}. We then plot the same for a randomly selected investor, who is not an opportunistic agent.  In addition to comparing the distributions of PDs, we also statistically confirmed that the opportunistic agents have different PDs in pre-announcement and non-announcement periods, and the difference in PDs as measured by Wasserstein distances is significantly greater than the non opportunistic agents (See \ref{app_add} for details).




We analyze the fitted distributions of PDs from observations during both the pre-announcement and non-announcement periods to identify differences in distribution parameters. This process begins with fitting the parametric GPD model, represented by the parameter set $\Theta = (\theta_1, \theta_2, \ldots, \theta_n)$. A crucial step in this process is determining the optimal number of parameters $n$, which we evaluate separately for homology groups $H_0$ and $H_1$. After systematically testing parameter counts from 1 to 4 using goodness-of-fit measures, we determine that the best fit requires 4 parameters for $H_0$ and 1 parameter for $H_1$.  Given two PDs with the parameters $\Theta_i = (\theta^i_1, \theta^i_2, \ldots, \theta^i_n) \mbox{ for } i=1,2$, we then assess statistical differences between the distributions using the test statistic $T$ defined previously.

\begin{figure}[!ht]
\begin{center}
		\includegraphics[scale=.75]{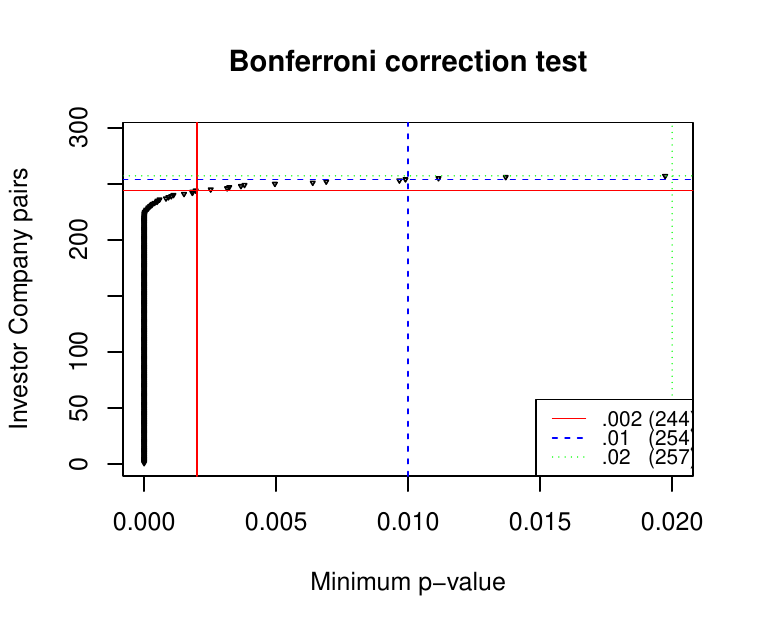}
    	
     \caption{\textbf{Minimum p-values with Bonferroni Correction t-Test Results:} The plot displays outcomes from t-tests examining the statistical differences in the distribution of persistence diagrams for opportunistic investor-company pairs between announcement and non-announcement periods. It highlights the minimum p-values across all the distribution model parameters, with horizontal lines indicating the threshold value \( \frac{\alpha}{m} \). A Green Line Represents \( \alpha=0.1 \), Blue Line Represents \( \alpha=0.05 \) and a Red Line Represents \( \alpha=0.01 \), where \( m=5 \) (4 for group \( H_0 \) and 1 for group \( H_1 \)). The corresponding lines represent the number of investors falling below the minimum p-value, as indicated in the legend. }\label{fig:minumum_p}
\end{center}
\end{figure}

First, we compare if the distribution of PDs  opportunistic agents are different between announcement and non-announcement periods. Our analysis revealed significant differences in the distribution of PDs between the periods. Given the identified 4 parameters for $H_0$ and 1 parameter for $H_1$, we applied Bonferroni correction to account for multiple comparisons. The results of these tests, including minimum p-values derived from the t-test, are shown in Figure \ref{fig:minumum_p}, highlighting the differences across all parameters for both homology groups. We analyzed 454 opportunistic agent-company pairs identified by our approach, but were able to fit the model to both pre-announcement and non-announcement period transaction data for 291 pairs, as the remaining lacked sufficient points in the PDs to fit the GPD. The test results reveal that for 244 of these pairs, we obtained p-values below 0.002, signifying statistically significant differences in the PD distributions between the two periods. Overall, we rejected the null hypothesis at a 90\% confidence level for 88\% of the investor-company pairs. Notably, for the 24 most traded companies, the null hypothesis was rejected with a 90\% confidence level in over 92\% of cases.

Second, we extended our investigation to determine whether the observed differences in the parameter estimates of the distributions for opportunistic agents are larger than those for non-opportunistic agents. For all opportunistic agent-company pairs we constructed a reference set as follows: for each agent-company pair $(A^s, C_j)$, where $A^s$ is the opportunistic agent and $C_j$ is the corresponding company, we identified a counterpart $(A^{ns}, C_j)$, where $A^{ns}$ is an agent who transacted with $C_j$ but was not labeled as an opportunistic. Let $T_s$ (defined in Eq. (\ref{eq:test})) represent the test statistic representing the difference in parameters of PDs for opportunistic agents during announcement and non-announcement periods and $T_r$ for the non-opportunistic counterparts. To assess whether $T_s$ is significantly greater than $T_r$, we conducted a hypothesis test with the null hypothesis $H_{null}: T_{r} \geq T_s$ and the alternative hypothesis $H_{alt}: T_{r} < T_s$. The intuition behind the alternative hypothesis is that opportunistic agents' trading behavior during the pre-announcement and non-announcement periods may differ more compared to non-opportunistic agents (the reference dataset). For robustness, we sampled the reference dataset 30 times and summary stats for p-values for the four parameters associated with homology group $H_0$ are presented in Table \ref{tab:pval}. Note that for homology group $H_1$, we were unable to draw any conclusions due to the limited number of one-dimensional features in the PDs. We observe that, across all parameters, the p-values are consistently less than 0.01, signifying substantial differences in the parameter estimates for the opportunistic agents during the pre-announcement and non-announcement periods than the non-opportunistic ones. Although the p-values for $\theta_0$ are relatively higher in some cases, the p-values for the remaining parameters remain significantly low, as reflected in the maximum p-values across these parameters. This pattern reinforces the conclusion that the observed differences in parameter estimates for opportunistic agents are statistically meaningful and higher than the non opportunistic agents.



\begin{table}[!ht]
  \centering
    \begin{tabular}{lrrrr}
    \toprule
          & \multicolumn{1}{c}{\textbf{$\theta_0$}} & \multicolumn{1}{c}{\textbf{$\theta_1$}} & \multicolumn{1}{c}{\textbf{$\theta_2$}} & \multicolumn{1}{c}{\textbf{$\theta_3$}} \\
    \midrule
     \textbf{Min.}  & 2.17E-03 & 3.82E-21 & 2.06E-26 & 1.21E-19 \\
    \textbf{Median} & 3.39E-03 & 5.00E-18 & 5.44E-22 & 6.09E-16 \\
    \textbf{Mean}  & 5.72E-03 & 7.35E-11 & 1.63E-10 & 1.53E-09 \\
    \textbf{Max.}  & 3.89E-02 & 1.12E-09 & 3.92E-09 & 4.60E-08 \\
    \bottomrule
    \end{tabular}%
     \caption{Statistical Analysis of p-Values on $H_{null}: T_{r} \geq T_s$: Evaluation of t-test results to assess whether the difference in parameter estimates of the Persistence Diagram distributions in the pre-announcement vs non-announcement periods for opportunistic agents is greater than that for other investors across all 30 reference datasets. The presented p-value summaries correspond to tests applied to all four parameters associated with group $H_0$. }
 
  \label{tab:pval}%
\end{table}%

\section{Discussion and Conclusion}

In this study, we introduced a Mapper-based topological data analysis approach that incorporates expert knowledge to identify opportunistic agents in financial markets---those who tend to exploit private information. This integration of expert knowledge into the Mapper framework is the key innovation of our approach. The method focuses on relevant features that enhance the accuracy of detecting patterns in the data. While we applied this method to financial markets, it is applicable to any contagion scenario, including the spread of diseases or rumors in networks.

First we validated our approach with Monte-Carlo experiment. Then we applied our approach to empirical trading data on actual investors and insiders. Focusing on company-investor pairs, we clustered agents who opportunistically use private information in trading. We found that the investors identified as opportunistic have higher connectivity to company insiders, from whom the insider information could originate. The statistical validation of our results using Gibbs distributions for persistence diagrams (GPD) further revealed significant differences in the trading behavior of opportunistic agents during pre-announcement versus other periods. Our finding is that opportunistic agents exhibited significantly different trading behaviors as measured by PDs of pre-announcement versus other periods, along with higher Wasserstein distances between the PDs during pre-announcement and non-announcement periods compared to the other agents. This finding supports the hypothesis that these agents are influenced by inside information. While computational challenges prevented us from applying the GPD method to all investor-company pairs, our targeted approach proved computationally feasible without sacrificing statistical rigor, making it an effective solution for large datasets.

Our Mapper-based approach offers potential beyond financial markets. It is suitable for analyzing various types of information cascades, such as the spread of diseases, social influence, or misinformation. To do so, requires defining appropriate domain-specific filter functions that capture the most relevant features of the analysed behavior. For example, in the case of epidemics, filter functions could be based on transmission rates, geographic location, or individual mobility patterns. In social influence networks, filter functions could be designed to measure the frequency and impact of interactions or the reach of messages. In misinformation networks, filter functions could focus on content virality or user engagement metrics. By tailoring the filter functions to the specific characteristics of the data, Mapper can potentially identify disease super-spreaders, key influencers, or sources of misinformation, making it a valuable tool for when agent type must be inferred from observable behaviors indirectly. Limitations of the approach remain. Applying GPD across all possible pairs would allow deeper analysis, but the computational demand is high. Future research could focus on computational aspects or approximations that maintain accuracy without sacrificing efficiency. The use of social network characteristics could enhance precision by directly modeling the flow of information between agents. 

In conclusion, the proposed approach is a significant advancement in the identification of symptomatic agents in hidden information contagions. Combining expert knowledge with topological data analysis and persistence homology, it offers an adaptable framework for tackling complex, multidimensional data. It is applicable across a wide range of domains. This positions the Mapper-based approach as a potentially transformative approach for challenging problems in data science and network analysis.


\section*{Data Availability }
The data that support the findings of this study are available from Euroclear Finland Ltd.; however, the raw data
cannot be distributed by the authors under the non-disclosure agreement signed with the data provider.

\section*{Acknowledgements}

The first author acknowledges support from the European Union's Horizon Europe programme under the Marie Skłodowska-Curie Actions (Grant Agreement No. 101150609, Project: HiddenTipChains).

\section*{Declaration of Generative AI and AI-assisted technologies in the writing process}

While preparing this work, the authors used GPT-4 to refine the language and enhance the clarity of the manuscript. After using this tool, the authors reviewed and edited the content as needed and take full responsibility for the content of the publication.

\section*{Author contributions statement}

\textbf{Anubha Goel:} Conceptualization, Data curation, Formal analysis, Investigation, Methodology, Visualization, Writing - original draft, Writing - review \& editing. \textbf{Henri Hansen:} Investigation, Visualization, Writing - review \& editing. \textbf{Juho Kanniainen:} Conceptualization, Investigation, Project administration, Resources, Software, Supervision, Writing - original draft, Writing - review \& editing.

\section*{Competing interests} The authors declare no competing interests. 
\bibliographystyle{elsarticle-harv} 
\bibliography{main}

\newpage
\appendix

\section{Monte-Carlo Experiment for Model Validation}
\label{app_monte}
\subsection*{Settings}

To validate the methods in our framework, we created synthetic data using the Independent Cascade (IC) model that establishes a ground truth for the type of each agent. This ground truth serves solely for validation purposes; the clustering of agents is conducted in an unsupervised manner using the method described in the Method section. To make the Monte Carlo experiment as realistic as possible, we utilize a graph that represents an actual insider network, within which we simulate random information cascades, using the independent cascade model. We use a larger dataset for the synthetic validation. In our dataset, there are a total of 1830 agents, of which 1596 have some transactions. 127 of these agents have no connections to the largest component of the graph, and are totally isolated; They serve as "sure false positives", i.e., they never have access to inside information. Below, we detail the underlying assumptions and the procedures employed in generating the synthetic dataset:

\begin{enumerate}
    \item \textbf{Graph Representation:} Let \( G = (V, E) \) be a directed graph where \( V \) denotes the set of nodes representing the agents, and \( E \) denotes the set of directed edges representing information links between agents. The total number of agents is represented by \( N = |V| \). 

    \item \textbf{Social Network Construction:} The social network structure among agents is drawn from 
    observed insider networks. Whenever two agents are insiders of the same company, we assume there exists a directed edge between them in both directions. We only consider the edges that connect agents to the largest component of this social graph, leaving some agents completely unconnected. 

    \item \textbf{Agent Classification by Information Sharing Probability:} Agents are randomly assigned to one of three classes. Each class has a different probability of sharing information, denoted by \( q_i \) for class \( i \),  \( i \in \{1, 2, 3\} \). The distributions from which \( q_i \)'s are drawn for different classes are specified in Table \ref{tab:simu_para}. Agents with no path to any other companies are considered \textit{unconnected agents} and are never informed. The \( q_i \)'s can be also understood as edge weights, \( w_{ij}= q_i\), between connected agents \( i \) and \( j \).

    \item \textbf{Agent Grouping and Trading Probability Modeling:} We model how agents behave if they receive information by assigning each into one of three distinct behavioral groups: Passive, Neutral, and Opportunistic agents. Each group has a different probability to take an advantage of information they have received, $p_{\text{Passive}}, p_{\text{Neutral}}, p_{\text{Opportunistic}}$, which are drawn from uniform distributions specified in Table \ref{tab:simu_para} .This addresses differences in reactivity to new information. 

    \item \textbf{Baseline Model:} The baseline model simulates agent behavior in the absence of market information, where all agents trade with a probability of \( p_{\text{Uninformed}} \), specified in Table \ref{tab:simu_para}. This model describe the transactions that are executed in the absence of private information received through the social connections.

    \item \textbf{Return distributions:}
    The returns of informed transaction  are assumed to follow a normal distribution, with a constant standard deviation mean. These returns, on average, are higher than the returns of transactions executed by the baseline model. The distributions are specified in Table \ref{tab:simu_para}.

    \item \textbf{Independent Cascade (IC) Model:} To simulate the diffusion of information, the IC model employs a probabilistic framework to propagate information from seed agents through the network. The following steps describe the IC model implementation:

    \begin{enumerate}
        \item Initialize the seed agents as the primary sources of information.
        \item Set the state of each agent as uninformed, except for the seed agents which are set as informed.
        \item Mark all edges as “not yet tried,” indicating that they have not been used to transmit information.
        \item For each iteration, identify the set of newly informed agents, starting with the seed agents, and attempt to inform their 1-hop neighbors:
            \begin{enumerate}
                \item For each informed agent, evaluate its outgoing edges to 1-hop neighbors.
                \item Generate a random variable \( r \sim U(0, 1) \). If the edge weight \( w_{ij} \) exceeds \( r \), the neighboring agent becomes informed, and the edge is marked as “tried.”
                \item Repeat this process iteratively until no new agents can be informed.
            \end{enumerate}
    \end{enumerate}

    This algorithm captures the stochastic nature of information spread within the network, allowing us to obtain the final state of each agent as either informed or uninformed.

\end{enumerate}

\begin{table}[htbp]
  \centering
  \caption{This table presents the division of agents into nine groups based on the probability of spreading information ($q_i,~i \in \{1,2,3\}$) and the probability of acting on the received information. Opportunistic agents exhibit a high probability ($p_{\text{Opportunistic}}$) of using the information for trading decisions, whereas passive agents have a low probability ($p_{\text{Passive}}$). Neutral agents exhibit a moderate probability ($p_{\text{Neutral}}$) of utilizing the information. Informed agents execute trades following a returns distribution with a higher mean compared to uninformed agents, with the same standard deviation. The values in the main matrix indicate the percentage of agents who spread the information with probability $q(\cdot)$ and act on the received information with probability $p(\cdot)$.}
\resizebox{1\textwidth}{!}{%
\begin{tabular}{lrlrr}
  \hline
        &       & \multicolumn{3}{c}{Probability to use information} \\
        \hline
    \multicolumn{2}{l}{\% Agents in Each Group} & $p_{\text{Passive}} \sim (0, 0.4)$ & \multicolumn{1}{l}{$p_{\text{Neutral}} \sim (0.3, 0.6)$} & \multicolumn{1}{l}{$p_{\text{Opportunistic}} \sim (0.5, 0.8)$} \\
    \midrule
    \multirow{3}[2]{*}{Prob. to distribute information} & $q_1 \sim U(0, 0.2)$ & \multicolumn{1}{r}{0.25} & 0.1   & 0.05 \\
          & $q_2 \sim U(0.2, 0.4)$ & \multicolumn{1}{r}{0.15} & 0.2   & 0.1 \\
          & $q_3 \sim U(0.4, 0.6)$ & \multicolumn{1}{r}{0.05} & 0.07  & 0.03 \\
    \midrule
          &       & \multicolumn{3}{c}{Return distribution} \\
          \hline
    \multicolumn{2}{l}{Informed agents} & \multicolumn{2}{l}{$\mathcal{N}(0.003, 0.0015)$} &  \\
    \multicolumn{2}{l}{Not informed agents} & \multicolumn{2}{l}{$\mathcal{N}(0.0005, 0.0015)$ or $\mathcal{N}(0.0015, 0.0015)$} &  \\
    \bottomrule
  \end{tabular}}%
  \label{tab:simu_para}%
\end{table}%

As a result, in each cascade or simulation, the information spreading starts from the seed company and spreads across the network based on each agent's probability of sharing information. Once informed, an agent then trades based on their designated probability of acting on the information.  Uninformed agents, regardless of their classification, trade according to a baseline model where no information prevails in the market. Using this setup, we simulate the investors' transactions and their profitability for the two periods i.e. the pre-announcement periods and non-announcement periods 1,000 times using the IC and baseline models respectively for each seed company, producing two \( N \times 1000 \) return matrices. These matrices are then concatenated to form a single \( N \times 2000 \) return matrix, which serves as the input for subsequent analyses. The aim is to effectively filter out agents who have received insider information and acted on it.

\subsection*{Methods for comparative analysis and Parameter Setting}

In evaluating the effectiveness of our proposed approach, we undertake a comparative analysis with two distinct categories of models: hard clustering techniques and unsupervised anomaly detection methods. The rationale behind this comparative approach is twofold: firstly, Mapper operates as a soft clustering technique, and we aim to gauge its performance against conventional hard clustering methods where we incorporate expert knowledge to identify the cluster corresponding to opportunistic agents. We use features extracted using the same filter functions as in the Mapper approach and compare the most commonly used clustering methods. This includes K-means \citep{macqueen1967some}, K-means++ \citep{arthur2007k}, Density-Based Spatial Clustering Application with Noise (DBSCAN) \citep{ester1996density}, and Hierarchical Clustering (Hclust) \citep{murtagh2012algorithms}. 

The exploitation of private information can be thought of as an anomaly in trading patterns. This is why we also check how the Mapper approach performs relative to commonly used unsupervised anomaly detection methods that do not incorporate expert knowledge. We use the One-Class Support Vector Machine (OCSVM), K-Nearest Neighbor (KNN), Isolation Forest (IForest), and Local Outlier Factor (LOF) \citep{chandola2009anomaly} methods for comparison. The benckmark methods were used as follows:
\begin{itemize}
    \item \textbf{DBSCAN:} We experimented by varying the minimum number of points (MinPts), in the range [30,50], and varied the epsilon (\(\epsilon\)) parameter within the range \([0.05, 1.5]\). After extensive testing, the optimal configuration was determined to be MinPts = 30 and \(\epsilon = 0.08\), which yielded the best results in terms of average silhouette scores.
    
    \item \textbf{Hierarchical Clustering:} We employed Ward's method for linkage, which minimizes the variance within clusters. The hierarchical tree was subsequently segmented into three groups to maintain alignment with the classification structure used in other methods.

    \item \textbf{One-Class SVM:} The One-Class Support Vector Machine model was exclusively trained on data from non-announcement periods to establish a baseline representation of normal behavior. Among the four tested kernels—sigmoid, polynomial, linear, and radial—the sigmoid kernel demonstrated superior performance with the lowest rate of false positives, making it the most reliable choice for detecting deviations indicative of suspicious trading activity.

    \item \textbf{KNN, LOF, and Random Forest:} Anomaly scores were assigned to each data point within our dataset based on these methods. Agents with anomaly scores exceeding the 88th percentile were flagged as Opportunistic agents, consistent with our Mapper approach, which effectively identified 208 (18\%) agents within the Opportunistic group.
\end{itemize}

For the Mapper approach, We set the number of bins to $|I|=3$, the percentage overlap $p=25\%$ and employ K-means clustering with 150 clusters with Euclidean distance as the measure of dissimilarity. To ensure a rigorous and unbiased comparison, a uniform post-clustering approach was adopted, whereby clusters were selected based on the weighted mean of standardized filter function values. Specifically, each filter function value was standardized using Z-score scaling, and a weighted average was computed with equal weights of 0.5 for each function value. This approach mitigates the influence of any single feature dominating the clustering outcome, thereby promoting a balanced assessment.

The number of clusters was set to three, aligning with our clustering of agents into three distinct groups. For the Mapper approach, a similar rationale guided our selection of three bins. The cluster with the highest mean weighted average value was designated as the group containing potentially Opportunistic agents. This methodological consistency across soft and hard clustering techniques enables a direct and unbiased evaluation of their efficacy in extracting meaningful clusters.

\subsection*{Results of the Monte Carlo Experiments}
Table \ref{simu_tab_2} reports the results of the Monte-Carlo experiments. It presents how accurately the opportunistic agents were identified with the proposed Mapper-based and benchmark methods. The results are based on two different baseline distributions for uninformed investors: $\mu = 0.0005$ and $\mu = 0.0015$ (see Table \ref{tab:simu_para}). The results show that when employing the Mapper algorithm, its ability to accurately identify members of opportunistic agents while minimizing false positives stands out, with a significant advantage over the other clustering methods. The Mapper approach is more cautious in labeling agents as opportunistic, which is often desirable in real-world applications where we aim to avoid falsely labeling passive individuals. We also report the F1-scores, which give equal weight to both precision and recall. However, depending on the context, different weightings can be applied to emphasize either precision or recall as needed.

In the context of hard clustering, one might initially anticipate the formation of distinct clusters based on the three predefined groups. However, the obtained results are rather surprising. In the case of K-means, K-means++, and hierarchical clustering (Hclust), the selected clusters encompass members from all three groups, including agents who were never exposed to insider information. This observation highlights the unreliability of these methods. DBSCAN does demonstrate a marked advantage compared to the other techniques. However, it struggles to effectively differentiate between neutral and opportunistic agents. 

\begin{table}[h!]
\centering
\caption{\label{simu_tab1}Performance of different models in identifying opportunistic agents under two different baseline model distributions, $\mu = 0.0005$ and $\mu = 0.0015$. Out of the 1830 total agents, 308 were made opportunistic. The table presents the predicted positives, actual true positives (TP), false positives (FP), true negatives (TN), false negatives (FN), precision, recall, and F1-score for each model.}
\begin{tabular}{@{}llllllllll@{}}
\toprule
\textbf{Model} & \textbf{Predicted} & \textbf{TP} & \textbf{FP} & \textbf{TN} & \textbf{FN} & 
\textbf{Precision} & \textbf{Recall} & \textbf{F1-score} \\
\textbf{} & \textbf{Positive} & \textbf{} & \textbf{} & \textbf{} & \textbf{} & 
\textbf{} & \textbf{} & \textbf{} \\
\midrule
\multicolumn{9}{c}{\textbf{Panel A: $\mu = 0.0005$}} \\
\midrule
Mapper approach & 212 & 208 & 4 & 1518 & 100 & \textbf{0.9811} & 0.6753 & \textbf{0.7994} \\
K-means         & 998 & 242 & 756 & 766  & 66  & 0.2425 & 0.7857 & 0.3703 \\
K-means++       & 1016 & 248 & 768 & 754  & 60  & 0.2441 & 0.8052 & 0.3752 \\
Hclust          & 1042 & 251 & 791 & 731  & 57  & 0.2409 & \textbf{0.8149} & 0.3735 \\
DBSCAN          & 360 & 206 & 154 & 1368 & 102 & 0.5722 & 0.6688 & 0.6177 \\
OCSVM           & 795 & 110 & 685 & 837  & 198 & 0.1382 & 0.3571 & 0.1987 \\
KNN             & 604 & 220 & 384 & 1138 & 88  & 0.3642 & 0.7143 & 0.4837 \\
LOF             & 1440 & 219 & 1221 & 301 & 89  & 0.1521 & 0.7110 & 0.2487 \\
IForest         & 695 & 224 & 471 & 1051 & 84  & 0.3223 & 0.7273 & 0.4474 \\
\midrule
\multicolumn{9}{c}{\textbf{Panel B: $\mu = 0.0015$}} \\
\midrule
Mapper approach & 204 & 199 & 5 & 1517 & 109 & \textbf{0.9755} & 0.6468 & \textbf{0.7771} \\
K-means         & 1022 & 245 & 777 & 745  & 63  & 0.2397 & 0.7955 & 0.3684 \\
K-means++       & 1025 & 246 & 779 & 743  & 62  & 0.2405 & 0.7987 & 0.3696 \\
Hclust          & 1056 & 248 & 808 & 714  & 60  & 0.2348 & 0.8052 & 0.3630 \\
DBSCAN          & 413 & 211 & 202 & 1320 & 97  & 0.5116 & 0.6851 & 0.5864 \\
OCSVM           & 1775 & 261 & 1514 & 8  & 47  & 0.1470 & \textbf{0.8474} & 0.2513 \\
KNN             & 578 & 216 & 362 & 1160 & 92  & 0.3737 & 0.7013 & 0.4873 \\
LOF             & 1283 & 225 & 1058 & 464 & 83  & 0.1753 & 0.7305 & 0.2837 \\
IForest         & 608 & 158 & 450 & 1072 & 150 & 0.2599 & 0.5129 & 0.3448 \\
\bottomrule
\end{tabular} \label{simu_tab_2}
\end{table}

As we increase the parameter $\mu$ for the baseline model distribution, making it more aligned with the behavior of informed investors, we observe a nuanced shift in the performance of different models. The Mapper approach, which initially exhibited a strong F1-score of 0.80 at $\mu = 0.0005$, shows a slight decrease to 0.78 at $\mu = 0.0015$. This minor decline suggests that while the Mapper approach remains robust, the challenge of distinguishing between informed and uninformed agents becomes more pronounced as their return distributions converge. Mapper maintains high precision (around 0.98) across both scenarios, indicating its reliability in minimizing false positives even when it correctly identifies fewer true positives due to the increased similarity in behavior.
DBSCAN, which initially had a relatively high F1-score of 0.62, experiences a slightly larger drop to 0.59 as $\mu$ increases. This is due to a rise in false positives from 154 to 202, highlighting its vulnerability to misidentification passive agents as opportunistic ones when the baseline profit increases. 
We also notice that traditional clustering methods like K-means and K-means++ struggle significantly with consistently low F1-scores around 0.37, regardless of $\mu$. Their high false positive rates and low precision suggest they are not well-suited for scenarios where informed and uninformed behaviors are closely matched.

Overall, while the Mapper approach performance slightly declines, it remains preferable due to its ability to maintain a balance between precision and recall. This conservative approach aligns well with the objective of accurately identifying opportunistic trading while avoiding excessive misidentification. The Mapper approach offers flexibility, as it allows for a straightforward analysis of the agents excluded in Step 3. Instead of removing these linked clusters, they can be considered for inclusion and further exploration, offering a more permissive and less conservative approach if desired. In contrast, models like DBSCAN and traditional clustering methods may require careful calibration or additional constraints to improve their effectiveness in similar contexts.

\section{Data pre-processing}\label{appdata}
We describe the various pre-processing procedures employed, which involve transforming raw data into a suitable format for further analysis. Let $N$ and $M$ denote the number of agents and companies, respectively. 

As in Ref. \citep{baltakiene2021identification}, for each agent \( i, i = 1, 2, \dots, N \) and her/his executed trade \( z \) in security \( k \), let \( Q_{i,k,z} \) be the quantity of shares traded, \( P_{i,k,z} \) the transaction price, and \( V_{i,k,z} = Q_{i,k,z} \times P_{i,k,z} \), the traded euro volume. To determine the profit and return from an investment made by agent $i$, for each transaction $z$ we define the $\Delta T$-day log return as
$$
r_{i,k,z}^{\Delta t} = \text{sign}(z) \times \log\left(\frac{P^{\Delta T}_{k,z}}{P_{i,k,z}}\right),
$$
where for security $k$, $P^{\Delta T}_{k,z}$ is the median transaction price on a day $\Delta T = 7$ calendar days after the transaction $z$ was executed, and $\text{sign}(z)$ is equal to $-1$ if $z$ was a sale, and $1$ if it was a purchase transaction. Here, $r_{i,k,z}^{\Delta t}$ captures the returns generated within a week after a trade. In cases where the price $P^{\Delta T}_{k,z}$ one week later is unavailable, we substitute it with the price observed on the next available trading day. The profit from the investment is then obtained by multiplying the return by the investment volume (in Euros). To calculate the realised returns for each day made by agent $i$ in the company $j$, we aggregate the profits from all investments made by the agent on that day in the given company we define the value-weighted average return for the agent $i$ in security $k$ as
$$
r_{i,k}^{\Delta t} = \frac{\sum_z r_{i,k,z}^{\Delta t} V_{i,k,z}}{\sum_z V_{i,k,z}}.
$$
This is necessary because an agent may conduct multiple trades in a single day involving different securities or the same securities of a single company.

Next, we elaborate on the methodology for calculating realised returns and profits during the pre-announcement period. For a specific company $j \in {1,\ldots,M}$ and a given time window $\delta=(t_j-|\delta|,t_j)$, where $|\delta| > 0$ (for example, a week with $|\delta| = 5$ or a month with $|\delta| = 20$, in this case, we consider one week), each agent $i$ is attributed a total return and net profit, summarizing their trading activities during that period. Here, $t_j$ denotes the announcement day when company $j$ makes public announcements, and $\delta$ represents the informed period, during which informed trading occurs in the securities of company $j$.

In scenarios where we have two announcements with overlapping periods, we handle the overlap by excluding the common period between the two announcements. For example, if an announcement is made for company $j$ on day $t$, and another announcement is made on day $t+3$, the informed period for the first announcement is defined as $(t-5, t)$, capturing the four trading days preceding the announcement day. This period represents the timeframe during which informed trading may have occurred based on the information disclosed regarding the first announcement. Similarly, for the second announcement, the informed period is $(t-2, t+3)$, covering the four trading days before the announcement day. This period reflects the interval during which informed trading could have taken place based on the information revealed in the second announcement. To ensure independence between the informed periods of the two announcements and avoid any potential overlap, we exclude the common period $[t-1]$ from the second informed period. Additionally, we define the non-announcement periods for each company as the trading days that fall outside the announcement period for that company, as mentioned above. These days do not include any announcements or periods of potential informed trading.

\section{Additional empirical results}\label{app_add}

\subsection{Empirical data Results with DBCSAN clustering}

The Table \ref{tab:dbscan_perfo} presents performance metrics for different companies and agents analyzed using DBSCAN clustering across four distinct cases. The aim is to differentiate between suspected opportunistic agents and others during announcement and non-announcement periods.

\begin{table}[ht!]
 \caption{Performance Metrics Comparison between opportunistic agents and other agents with DBCSAN. The table presents results for different cases of analysis. Case 1 encompasses all companies, while cases 2, 3, and 4 involve subsets of 24, 18, and 11 companies, respectively, filtered based on minimum total transactions made by agents in those companies (5,000, 6,000, and 7,000 respectively).}\label{tab:dbscan_perfo}
\resizebox{\textwidth}{!}{%
  \centering
  \begin{tabular}{lrrrrrrrr}
    \toprule
    \textbf{ } & \multicolumn{2}{c}{\textbf{All data}} & \multicolumn{2}{c}{\textbf{Data for 24 most}} & \multicolumn{2}{c}{\textbf{Data for 18 most}} & \multicolumn{2}{c}{\textbf{Data for 11 most}} \\
    \textbf{ } & \multicolumn{2}{c}{\textbf{}} & \multicolumn{2}{c}{\textbf{ traded companies}} & \multicolumn{2}{c}{\textbf{ traded companies}} & \multicolumn{2}{c}{\textbf{ traded companies}} \\
    
    \midrule
    \textbf{Companies} & \multicolumn{2}{c}{119} & \multicolumn{2}{c}{24} & \multicolumn{2}{c}{18} & \multicolumn{2}{c}{11} \\
    \textbf{Number of agents} & \multicolumn{2}{c}{1,586} & \multicolumn{2}{c}{1,217} & \multicolumn{2}{c}{1,179} & \multicolumn{2}{c}{1,112} \\
    \textbf{Number of agent-company pairs} & \multicolumn{2}{c}{15,668} & \multicolumn{2}{c}{8,311} & \multicolumn{2}{c}{7,169} & \multicolumn{2}{c}{4,532} \\
    \textbf{Minimum number of transactions} & \multicolumn{2}{c}{59} & \multicolumn{2}{c}{5,000} & \multicolumn{2}{c}{6,000} & \multicolumn{2}{c}{7,000} \\
    
    \midrule
   \textbf{
   } &\multicolumn{1}{c}{\textbf{Opportunistic}} & \multicolumn{1}{c}{\textbf{Others}} & \multicolumn{1}{c}{\textbf{Opportunistic}} & \multicolumn{1}{c}{\textbf{Others}} & \multicolumn{1}{c}{\textbf{Opportunistic}} & \multicolumn{1}{c}{\textbf{Others}} & \multicolumn{1}{c}{\textbf{Opportunistic}} & \multicolumn{1}{c}{\textbf{Others}} \\
   \midrule

    \textbf{Agents} & 1268  & 318   & 913   & 333   & 867   & 319   & 856   & 318 \\
     \multicolumn{1}{p{5cm}} {   \textbf{Fraction of euro volume in announcement period} } & 53\%  & 16\%  & 34\%  & 29\%  & 26\%  & 28\%  & 26\%  & 27\% \\
     \multicolumn{1}{p{5cm}} { \textbf{Profit in announcement period per agent} }& -149.0 & -45.3 & -245.9 & 2.2   & 423.0 & 3.4   & 458.2 & 3.0 \\
     \multicolumn{1}{p{5cm}} {   \textbf{Profit in non-announcement period per agent} }& 2987.5 & 1044.4 & 830.8 & -70.2 & 478.5 & -61.5 & 563.1 & -61.4 \\
    \multicolumn{1}{p{5cm}} {\textbf{Fraction of profitable transactions in announcement period vs all profitable transactions}} & 29\%  & 32\%  & 31\%  & 34\%  & 31\%  & 33\%  & 31\%  & 32\% \\
    \multicolumn{1}{p{5cm}}{ \textbf{Fraction of unprofitable transactions in announcement period vs all unprofitable transactions} } & 27\%  & 30\%  & 30\%  & 34\%  & 30\%  & 33\%  & 29\%  & 34\% \\
    \bottomrule
  \end{tabular}%
  }
\end{table}%

The number of companies varies across the cases, starting with 119 in Case 1 and decreasing to 11 in Case 4 based on Minimum transaction thresholds. The DBSCAN clustering results show a larger proportion of agents identified with suspicious trading behavior in each case, suggesting that many agents are flagged as opportunistic.

One of the key observations is that opportunistic agents consistently control a larger fraction of euro volume during announcement periods compared to Others. For instance, in Case 1, opportunistic agents handle 53\% of the volume, while Others only manage 16\%. However, the data reveal a significant flaw: despite being suspected of opportunistic behavior, agents consistently show negative profits during announcement periods across all cases. For example, in Case 1, opportunistic agents average a loss of -149.0 euros, while Others incur a smaller loss of -45.3 euros. This trend of negative profits continues in the subsequent cases, challenging the notion that  opportunistic agents are exploiting market-sensitive information for gains.

Furthermore,  opportunistic agents show higher profits in non-announcement periods, which contradicts the expected behavior of opportunistic agents who would typically focus on profiting during high-information periods/ announcements periods. For instance, in Case 1, opportunistic agents earn an average profit of 2987.5 euros in non-announcement periods, compared to 1044.4 euros for Others. Additionally, the fractions of profitable and unprofitable transactions during announcement periods for opportunistic agents are nearly identical or smaller than others, indicating that suspected agents are not particularly better or worse at timing their trades compared to others.

Overall, the data do not support the identification of most agents as opportunistic, as their actual trading results during announcement periods do not align with expected opportunistic behavior. The clustering approach, while insightful, appears to misidentify agents who may not be trading advantageously around announcements. Instead, their profits predominantly arise outside these periods, suggesting that the DBSCAN method may need refinement to accurately identify truly opportunistic agents. This misalignment between expected and observed behaviors highlights the inefficiency of DBSCAN approach.

\subsection{Attributes of agents identified by the Mapper approach}
\begin{table}[ht]
\centering
\begin{minipage}{0.4\textwidth}
\centering
\caption{Percentage of agents identified as opportunistic across different regions, age groups, gender, and experience levels}
\begin{tabular}{lr}
\toprule
\textbf{Metric} & \textbf{Opportunistic}  \\
\midrule
\multicolumn{2}{c}{\textbf{Region Counts}} \\
\midrule
Central-Finland & 31.82\% \\
Eastern-Finland & 23.81\% \\
Eastern-Tavastia & 16.00\% \\
Helsinki & 17.08\% \\
Northern-Finland & 15.15\% \\
Northern-Savonia & 10.00\% \\
Ostrobothnia & 14.08\% \\
Rest-Uusimaa & 14.12\% \\
South-East & 15.79\% \\
South-West & 16.04\% \\
Western-Tavastia & 10.34\% \\
\midrule
\multicolumn{2}{c}{\textbf{Age Counts}} \\
\midrule
$<$30 & 9.09\% \\
31-40 & 14.18\% \\
41-50 & 14.51\% \\
51-60 & 11.72\% \\
61 and above & 23.14\% \\
\midrule
\multicolumn{2}{c}{\textbf{Gender Counts}} \\
\midrule
Number of Males & 15.43\% \\
Number of Females & 11.94\% \\
\midrule
\multicolumn{2}{c}{\textbf{Experience Range}} \\
\midrule
0-1 year & 3.66\% \\
1-2 years & 4.46\% \\
2-3 years & 5.45\% \\
3-4 years & 7.34\% \\
4+ years & 26.12\% \\
\bottomrule
\end{tabular} \label{att1}
\end{minipage}%
\hfill
\begin{minipage}{0.5\textwidth}
\centering
\caption{Distribution of agents identified as opportunistic and non-opportunistic according to region, age group, gender, and experience level}
\begin{tabular}{lrr}
\toprule
\textbf{Metric} & \textbf{Opportunistic} & \textbf{Others} \\
\midrule
\multicolumn{3}{c}{\textbf{Region Counts}} \\
\midrule
Central-Finland & 2.93\% & 1.23\% \\
Eastern-Finland & 2.09\% & 1.32\% \\
Eastern-Tavastia & 1.67\% & 1.73\% \\
Helsinki & 69.04\% & 65.87\% \\
Northern-Finland & 2.09\% & 2.30\% \\
Northern-Savonia & 0.84\% & 1.48\% \\
Ostrobothnia & 4.18\% & 5.02\% \\
Rest-Uusimaa & 5.02\% & 6.00\% \\
South-East & 1.26\% & 1.32\% \\
South-West & 7.11\% & 7.32\% \\
Western-Tavastia & 3.77\% & 6.41\% \\
\midrule
\multicolumn{3}{c}{\textbf{Age Counts}} \\
\midrule
$<$30 & 2.46\% & 4.33\% \\
31-40 & 9.85\% & 10.47\% \\
41-50 & 35.96\% & 37.20\% \\
51-60 & 24.14\% & 31.92\% \\
61 and above & 27.59\% & 16.09\% \\
\midrule
\multicolumn{3}{c}{\textbf{Gender Counts}} \\
\midrule
Number of Males & 82.38\% & 77.64\% \\
Number of Females & 17.62\% & 22.36\% \\
\midrule
\multicolumn{3}{c}{\textbf{Experience Range}} \\
\midrule
0-1 year & 1.18\% & 6.00\% \\
1-2 years & 2.75\% & 11.39\% \\
2-3 years & 4.31\% & 14.50\% \\
3-4 years & 9.41\% & 23.01\% \\
4+ years & 82.35\% & 45.10\% \\
\bottomrule
\end{tabular}\label{att2}
\end{minipage}
\end{table}

We expanded the study by addressing a key question in insider trading: "Do insiders who disseminate information share common attributes such as age, postal code, organizational role, experience, portfolio size, trading activity, and more?" This question is intrinsically tied to the concept of homophily, as explored in studies such as \cite{kossinets2009origins}.
Table \ref{att1} lists the percentage of agents that were profiled as opportunistic agents, by the Mapper approach, according to region, age group, gender, and experience level.  
Table \ref{att2} presets how opportunistic and non-opportunistic agents were distributed according to different regions, age groups, and experience range. We see from Table \ref{att1}, the highest proportion of opportunistic agents is found in Central Finland (31.82\%) and Eastern Finland (23.81\%), with a noticeable presence in Helsinki (17.08\%). When considering age groups, the data reveals that opportunistic agents are more common in the age group 61 and above (23.14\%), suggesting that older investors are more inclined towards opportunistic investment strategies. Gender distribution shows a males (15.43\%) are slightly more likely to be profiled as opportunistic than females (11.94\%). Additionally agents with 4+ years of experience are most likely (26.12\%) to be profiled as opportunistic. 

From Table \ref{att2}  shows that most of the agents, 69.04\% of opportunistic and 65.87\% of others, are from Helsinki region. This is not surpising given the importance of the capital region. The distribution over the other regions reflects mostly the economic importance of the said regions. In terms of distribution by age, the largest group of agents is the age group 41--50, but only in the 61+ age group do we see a marked overrepresentation of opportunistic investors. When looked at the experience range, the contrast becomes even starker. 

Together, these findings underscore the importance of investor experience and regional factors in shaping investment strategies. In particular, the prevalence of opportunistic behavior in regions like Helsinki suggests that access to markets, financial resources, and information may play a significant role. 
Central Finland seems anomalous in this sense, as it has a markedly higher percentage of opportunistic agents compared with others. The data indicates that older, more experienced investors tend to adopt opportunistic strategies more frequently, reinforcing the influence of accumulated market knowledge and experience on investment decisions. Additionally, the gender imbalance observed in both categories highlights a potential area for further exploration regarding the role of gender in investment decision-making.

\subsection{Visualization of Topological Summaries }

\begin{figure}[ht!]
\begin{center}
  \subfigure[From left to right: Projected Persistence Diagrams constructed using superlevel set filtration for two randomly selected opportunistic agents during the announcement and non-announcement periods, followed by Persistence Barcodes representing the persistence of topological features in each period. Black dots and red triangles denote the persistence of zero-dimensional and one-dimensional features, respectively.]{%
    \begin{minipage}{0.65\textwidth}
      \includegraphics[width=\textwidth]{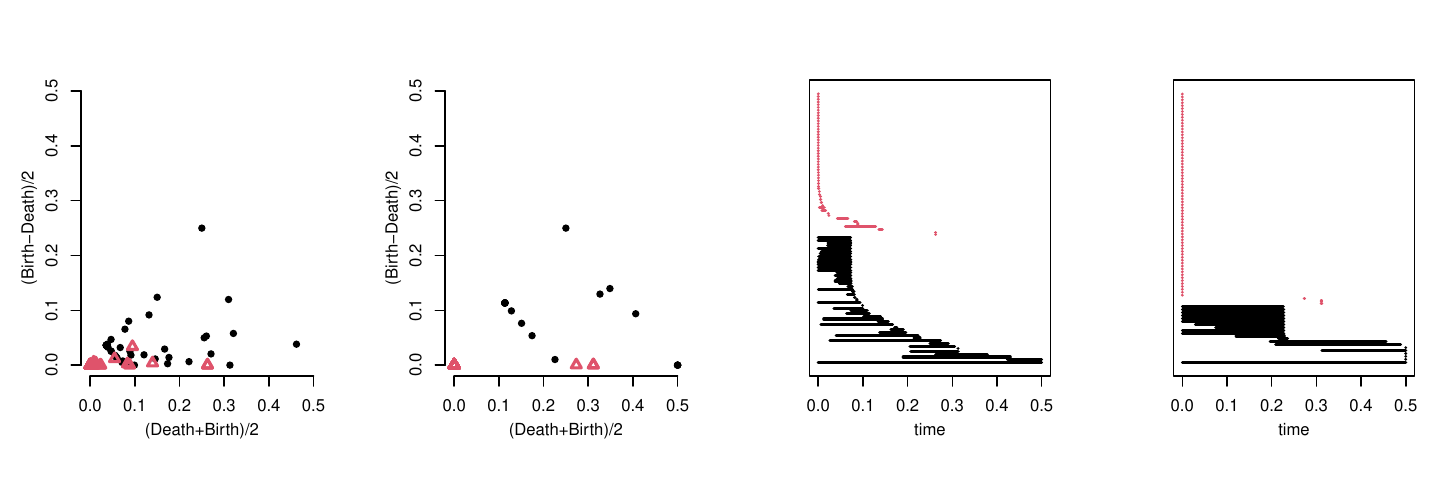}
      \includegraphics[width=\textwidth]{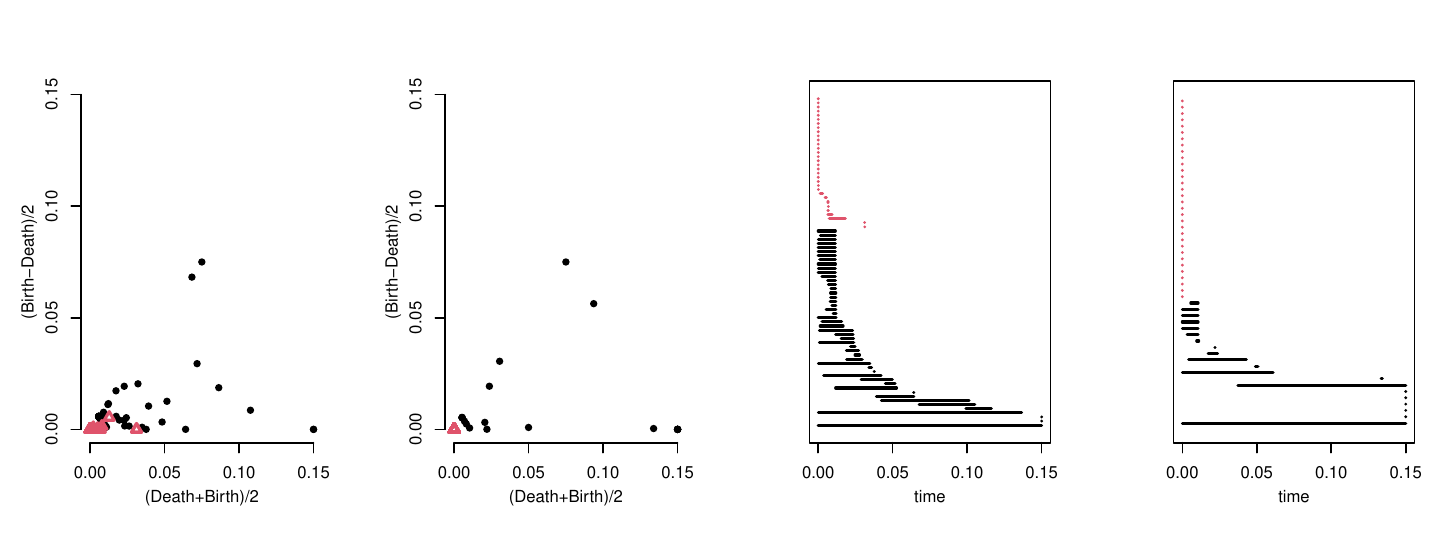}
    \end{minipage}
    \label{homology}}

  \hspace{0.5cm} 

  \subfigure[From left to right: Projected Persistence Diagrams constructed using superlevel set filtration for two randomly selected non-opportunistic agents during the announcement and non-announcement periods. Black dots and red triangles denote the persistence of zero-dimensional and one-dimensional features, respectively.]{%
    \begin{minipage}{0.66\textwidth}
      \includegraphics[width=\textwidth]{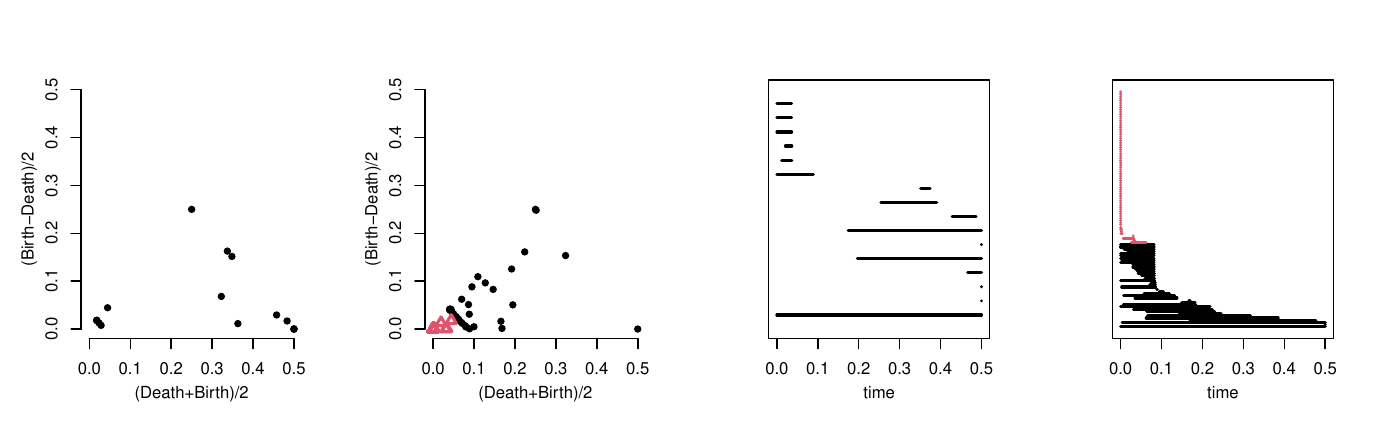}
      \includegraphics[width=\textwidth]{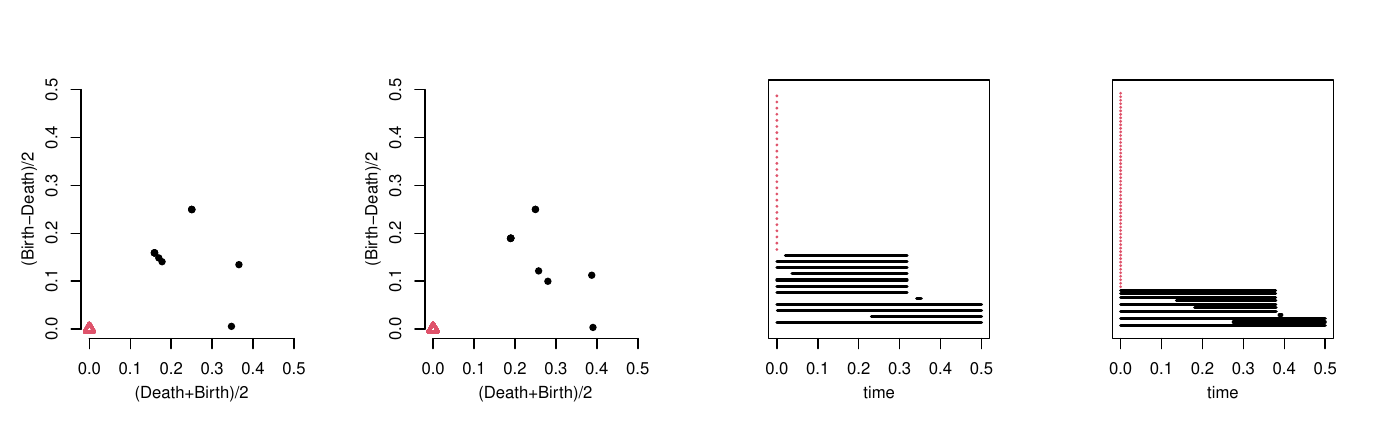}
    \end{minipage}
    \label{homology_reference}}
\end{center}
\caption{Illustration of Persistence Diagrams and Persistence Barcodes for two opportunistic agents and two non-opportunistic agents.}
\label{fig2:pds}
\end{figure}

Figure \ref{fig2:pds} presents a set of topological summaries—specifically, Persistence Diagrams and Persistence Barcodes—that provide insights into the behaviors of two distinct types of agents: opportunistic agents and passive (non-opportunistic) agents. These summaries are derived from randomly selected agents within each category, offering a comparative analysis of their trading patterns.

Persistence Diagrams (left two plots in each row) illustrate the birth and death of topological features in the data. In these diagrams, black dots represent zero-dimensional features (connected components), and red triangles represent one-dimensional features (loops). The x-axis denotes the birth time (when a feature emerges), while the y-axis indicates persistence (the duration before the feature disappears). Persistence Barcodes (right two plots in each row) visually summarize the lifespan of these topological features, where each horizontal line corresponds to a feature, and its length indicates the feature’s persistence.

The Persistence Diagrams and Barcodes for opportunistic agents (Figure \ref{homology}) demonstrate a markedly higher persistence of topological features during the pre-announcement period compared to the non-announcement period. This increased persistence suggests that opportunistic agents engage in more prolonged and strategically complex trading behaviors during periods of information asymmetry. Their ability to sustain these intricate patterns indicates that they may be exploiting market inefficiencies or leveraging non-public information to optimize profits prior to public announcements. This behavior contrasts sharply with the non-announcement period, where their trading strategies appear less pronounced, underscoring the opportunistic nature of their activities.

In contrast, the Persistence Diagrams and Barcodes for passive agents (Figure \ref{homology_reference}) reveal two distinct behavioral patterns. One agent exhibits lower persistence during the pre-announcement period compared to the non-announcement period, possibly indicating a more cautious or reactive approach, avoiding significant market moves in anticipation of announcements. Conversely, the other agent shows consistently high persistence across both periods, suggesting a stable and systematic trading approach that remains unaffected by short-term information asymmetries. This consistent behavior reflects a disciplined trading strategy that contrasts with the more opportunistic patterns observed in other agents.
These two distinct patterns among passive agents highlight the utility of TDA in differentiating agent behaviors. 

\subsection{Persistence diagram comparison for the opportunistic agents as identified by the Mapper approach}

To take advantage of the topological information and topological properties derived by persistent homology, one must be able to compare persistence diagrams, i.e. endow the space of persistence diagrams with a metric structure. The similarity between the two persistence diagrams is commonly measured by the Wasserstein distance. The $p$ Wasserstein distance \citep{edelsbrunner2010computational} for $p\geq 1$, between two persistence diagrams $D_{ {X}}$ and $D_{ {Y}}$ corresponding to data sets $X$ and $Y$ is defined as 
 \begin{equation*}
 WD_p(D_{ {X}},D_{ {Y}})=\left( \inf_{\gamma} \sum_{a \in {D_{ {X}}}} ||a-\gamma(a) ||_{\infty}^p \right)^{\frac{1}{p}}.      
 \end{equation*}

where $\gamma$ ranges over all bijections from the points in $D_{ {X}}$ to the points in $D_{ {Y}}$. One can assume that a persistence diagram is a countable multi-set of points in $\mathbb{R}^2$ along with the diagonal where each point on the diagonal has infinite multiplicity. In this way the cardinalities of $D_{ {X}}$ and $D_{ {Y}}$ are considered to be equal. The Wasserstein distance provides a natural way to quantify the dissimilarity between two diagrams, thereby measuring the difference in the topological features of the underlying data sets. It offers robustness and stability by ensuring small data perturbations result in proportionally small changes in the distance, making it reliable for noisy data \citep{cohen2010lipschitz}.


We compare the persistence diagrams of a given group (opportunistic agents/other agents) across both pre-announcement and non-announcement periods by computing the Wasserstein distance. 
Two significant observations become apparent when using the Wasserstein distances between the persistence diagrams of the pre-announcement and non-announcement periods. First, we observe that for opportunistic agents, the Wasserstein distance significantly deviates from zero. This observation was statistically validated using a two-tailed t-test, which indicated that the distances were markedly different from zero at a 99\% confidence level, with p-values well below $2.2 \times 10^{-16}$ for both groups \(H_0\) and \(H_1\). This result implies that opportunistic agents traded differently between pre- and non-announcement periods in terms of the topological properties of the data.

Secondly, we ask whether the Wasserstein distance is greater for opportunistic agents than for other agents, anticipating that the distance would be greater for opportunistic agents because private information should play a bigger role in their trading. We first compute the Wasserstein distance for reference datasets constructed as follows: for each agent-company pair $(A^s, C_j)$, where $A^s$ is the opportunistic agent and $C_j$ is the corresponding company, we identified a counterpart $(A^{ns}, C_j)$, where $A^{ns}$ is an agent who transacted with $C_j$ but was not labeled as an opportunistic. We then conducted a right-tailed t-test with the null hypothesis $W_s - W_r = 0$, where $W_s$ is the Wasserstein distance for the opportunistic agent and $W_s$ is the Wasserstein distance for the agent not profiled as opportunistic. The alternative hypothesis is $W_s - W_r > 0$, (we anticipate a greater distance for opportunistic agents). For group $H_0$, the calculated p-value was below $2.2 \times 10^{-16}$, leading to the rejection of the null hypothesis. This result indicates that opportunistic agents do not only trade abnormally but also in an exceptionally different manner during the pre-announcement periods compared to other investors. On the other hand, for group $H_1$, there was insufficient evidence to reject the null hypothesis i.e. we could not statistically confirm that the Wasserstein distances are different. The reason behind this is the lack of one-dimensional features in the transaction data space due to the sparse nature of the data. This is clearly visualized in the Supplementary Material, where we plot the persistence diagrams of randomly selected agents. The diagrams predominantly consist of birth death pairs of one-dimensional features.

We conducted two robustness checks of these results. First, we randomly sampled the reference investors 30 times and computed the average Wasserstein distances. The p-value remained below $2.2 \times 10^{-16}$, supporting our earlier rejection of the null hypothesis and confirming the robustness of our approach. Second, the same tests were performed using the bottleneck distance, a special case of the Wasserstein distance with $p=\infty$, yielding identical results.

\end{document}